\documentclass[aip,jcp,reprint,longbibliography]{revtex4-1}

\usepackage[usenames,dvipsnames]{color}
\usepackage{amssymb}
\usepackage{amsmath}
\usepackage{hyperref}
\usepackage{graphicx}
\usepackage{subfigure}
\usepackage{xspace}

\renewcommand{\vec}[1]{\mathbf{#1}}
\newcommand{\vu}{{\vec u}}
\newcommand{\vr}{{\vec r}}
\newcommand{\DG}{\Delta G}
\newcommand{\DGij}{\Delta G_{ij}}
\newcommand{\dee}{{\mathrm{d}}}

\newcommand{\cZ}{{\cal Z}}
\newcommand{\cS}{{\cal S}}

\newcommand{\cnf}{{\text{(cnf)}}}
\newcommand{\hyb}{{\text{hyb}}}
\newcommand{\att}{{\text{att}}}
\newcommand{\rep}{{\text{rep}}}
\newcommand{\kT}{k_{\text{B}} T}

\newcommand{\from}{\,\leftarrow\,}

\DeclareMathOperator{\erf}{erf}

\pacs{64.10.+h,31.15.xv}

\date{\today}

\begin{document}

\title{A general theory of DNA-mediated and other valence-limited colloidal
  interactions}
\author{Patrick Varilly, Stefano Angioletti-Uberti, Bortolo M.\ Mognetti and
  Daan Frenkel}
\address{Department of Chemistry, University of Cambridge, Lensfield
  Road, CB2 1EW Cambridge, UK}

\begin{abstract}
We present a general theory for predicting the interaction potentials
between DNA-coated colloids, and more broadly, any particles that interact
via valence-limited ligand-receptor binding.  Our theory correctly
incorporates the configurational and combinatorial entropic factors that
play a key role in valence-limited interactions.  By rigorously enforcing
self-consistency, it achieves near-quantitative accuracy with respect to
detailed Monte Carlo calculations.  With suitable approximations and in
particular geometries, our theory reduces to previous successful treatments,
which are now united in a common and extensible framework.  We expect our
tools to be useful to other researchers investigating ligand-mediated
interactions.  A complete and well-documented Python implementation is
freely available at \url{http://github.com/patvarilly/DNACC}.
\end{abstract}

\maketitle


\section{Introduction}

The theoretical study of the statistical behavior of soft matter systems
generally begins with a description of the interaction between their
constituent particles.  One exciting system that has emerged in recent years
is that of DNA-coated colloids (DNACCs), where interactions between colloids
are driven by partial hybridization of complementary ssDNA strands grafted
on their surface~\cite{MirkinEtAl:1996, AlivisatosEtAl:1996,
  BiancanielloKimCrocker:2005, GeertsEiser:2010}.  To date, DNACCs have been
used mainly for building highly sensitive biomedical
probes\cite{RosiMirkin:2005} and to assemble binary nanoparticle crystals of
numerous morphologies\cite{NykypanchukEtAl:2008, ParkEtAl:2008,
  MayeEtAl:2010, MacfarlaneEtAl:2011}.

The aim of this paper is to present a general theory of DNACC interactions
that can be applied to any DNACC system, using only mild assumptions that
are valid for common DNACCs.  A key feature of our approach is that once the
binding statistics of an arbitrary pair of DNA strands is known, all
combinatorial and competitive effects can be computed separately.
Disentangling these two aspects is what gives our theory its generality.
Our efforts build on past theoretical attempts to understand DNACC
experiments in terms of the statistical mechanics of bond formation
\cite{BiancanielloKimCrocker:2005, LicataTkachenko:2006,
  NykypanchukEtAl:2007, DreyfusEtAl:2009, DreyfusEtAl:2010,
  LeunissenEtAl:2010, LeunissenFrenkel:2011, MognettiLeunissenFrenkel:2012,
  RogersCrocker:2011, MognettiEtAl:2012}, which have all partially succeeded
in modeling various experimental setups, differing for instance in coating
densities, nucleotide sequences, strand lengths, stiffness and colloid
sizes.  To date, however, a rigorous and general solution to the problem has
not been presented\cite{RogersCrocker:2011, MognettiEtAl:2012,
  RogersCrocker:2012}.

A salient feature of DNACCs is that their interaction is mediated by
surface-grafted ligands, each of which can form a single bond at a time.
Owing to this restriction, we say that such interactions are
``valence-limited''~\cite{ZaccarelliEtAl:2006}.  While we phrase our theory
in the language of DNACCs, the treatment is equally applicable to any
valence-limited interaction, including multi-valent
particles for drug delivery, viruses and
cells~\cite{Martinez-VeracoecheaFrenkel:2011,
  KrishnamurthyEstroffWhitesides:2006, BadjicEtAl:2005,
  MammenChoiWhitesides:1998}.

The rest of the paper is structured as follows.  In
Section~\ref{sec:theory}, we present a statistical mechanical derivation of
our theory in its most general form, followed by a specialization
to DNACCs and a computationally simpler mean-field approximation.  In
Section~\ref{sec:results}, we compare the results of our theory with those
from more accurate and computationally expensive Monte Carlo simulations.
We have chosen as examples specific DNACC systems that have been presented
previously~\cite{LeunissenFrenkel:2011, MognettiLeunissenFrenkel:2012,
  RogersCrocker:2011, Angioletti-UbertiMognettiFrenkel:2012}, as well as two
novel ones, to show that they can all be described using a single, unifying
framework.  In Section~\ref{sec:summary}, we summarize our approach and
reproduce the essential equations for reference purposes. Finally, in
Section~\ref{sec:conclusions}, we discuss possible applications of our
theory, both for DNACCs and for more general valence-limited interactions.

In Appendix~\ref{app:entropic}, we discuss an important technical aspect of
our theory, namely, how to calculate the entropic cost of binding for two
general polymeric ligands.  We also summarize the results for short, stiff
rods (which model short dsDNA tethers common in micron-sized
DNACCs~\cite{BiancanielloKimCrocker:2005, LeunissenEtAl:2010,
  GeertsEiser:2010}) and ideal chains (as used in
Refs.~\onlinecite{RogersCrocker:2011, MognettiEtAl:2012}).
Appendix~\ref{app:lce} explains how our theory is connected to the Local
Chemical Equilibrium treatment recently introduced by Rogers and
Crocker~\cite{RogersCrocker:2012} (itself an improvement of the work of
Biancaniello, Kim and Crocker in
Ref.~\onlinecite{BiancanielloKimCrocker:2005}). We show how a physically
motivated and simple change to the equations used by these authors corrects
their theory's principal shortcoming, and that the corrected version reduces
to a specialization of our theory.  Finally, Appendix~\ref{app:saddlepoint}
presents an alternate derivation of the mean-field approximation to our
theory using a saddle-point analysis, making explicit the connection between
our theory and earlier successful but more specialized
approaches~\cite{MognettiLeunissenFrenkel:2012,
  Angioletti-UbertiMognettiFrenkel:2012}.

Although this paper explains our approach in full detail, we emphasize that
the final results are simple and can be used widely by other researchers to
calculate pair potentials and binding configurations of general DNACCs.  To
this end, we have made freely available a complete and well-documented
Python implementation of the theory, which can be found at
\url{http://github.com/patvarilly/DNACC}.


\section{Theoretical Model}
\label{sec:theory}

Under experimentally relevant conditions, DNA-coated colloids interact
through entropic repulsion and hybridization-mediated
attraction\cite{BiancanielloKimCrocker:2005, DreyfusEtAl:2009,
  LeunissenFrenkel:2011}. Whereas the repulsion is simple to estimate for
common geometries, as shown in Appendix~\ref{app:entropic}, calculating the
attraction is more complicated.  The key subtlety is that the attraction is
mediated by valence-limited binding, i.e., each DNA strand can bind at most
one partner at a time~\footnote{Although one can design a strand that binds
  more than one partner simultaneously, doing so is uncommon for DNACCs,
  since the entropic cost of bringing three or more strands together when
  they are grafted onto colloids (as opposed to in solution) is
  prohibitive.}.  In this section, we first present a general treatment of
this attraction for generic valence-limited ligands.  Then, we specialize
our treatment to DNACCs.  Finally, we construct a mean-field approximation
that is suitable for geometries with translational symmetry.

\subsection{Statistical Mechanics of valence-limited interactions}
\label{sec:theory:statmech}

\begin{figure}
\begin{center}
\includegraphics{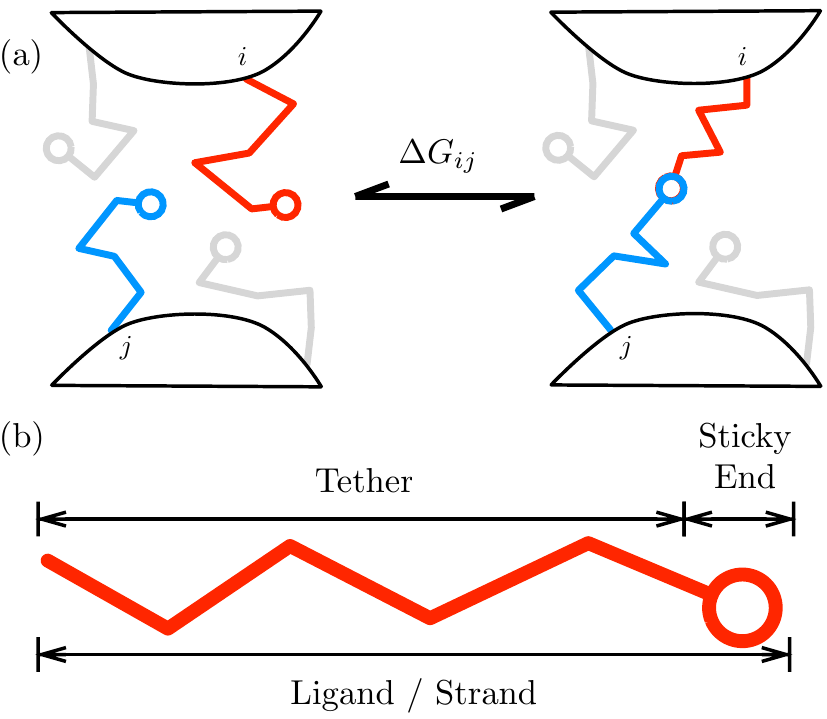}
\end{center}
\caption{\label{fig:system}(a) Schematic of a DNACC system.  Any pair of
  ligands, $i$~and~$j$, can form a bond at free energy cost $\Delta G_{ij}$,
  which is approximately independent of the presence of the other ligands.
  (b) Anatomy of a typical DNACC ligand, consisting of a long tether capped
  by a point-like sticky end.  Throughout, we use the terms ligand and
  strand interchangeably.}
\end{figure}

Consider a system of $N$ generalized ligands, depicted in
Figure~\ref{fig:system}(a). At any given time, an individual ligand $i$ can
bind at most one other ligand~$j$, with a resulting free energy
change~$\DGij$. If $i$~and~$j$ cannot bind, either for geometric or chemical
reasons, we set $\DGij$ to infinity.  Since $i$ cannot bind to itself, we
also set $\DG_{ii}$ to infinity for all~$i$.  The set of all $\DGij$ values
completely characterizes the relevant chemical detail of the system.  We
explicitly assume that $\DGij$ is independent of the presence of other
ligands, which is accurate when the density of ligands is low enough that two
ligands not bound to each other can be thought of as not interacting at all.
This condition holds for the grafting densities and salt concentrations
typically used in DNACC experiments with micron-sized
colloids~\cite{BiancanielloKimCrocker:2005, ValignatEtAl:2005,
  LeunissenEtAl:2010, DreyfusEtAl:2009}.

Let $\phi$ be a binding configuration of the system, i.e. a list of bound
ligand pairs $(i,j)$.  Taking the state where no bonds are formed as a
reference state, the free energy of $\phi$ is the sum of the binding free
energies of each bond.  Summing over all $\phi$, we obtain the exact partition
function of the system, $Z$, relative to that of the reference state, $Z_0$:
\begin{equation}
\frac{Z}{Z_0} = \sum_{\{\phi\}} \prod_{(i,j)\in \phi} e^{-{\beta \DGij}}.
\label{eqn:Z}
\end{equation}

For a particular ligand~$i$, the terms in the sum above can be classified
according to the binding partner of~$i$, if any.  Let $Z_{-i}$ be the
partition function for a system where $i$ is unbound, and let $Z_{-i,-j}$
likewise be the partition function for a system where both $i$~and~$j$ are
unbound.  The sum of the terms in Eq.~\eqref{eqn:Z} where $i$ is unbound is
$Z_{-i}/Z_0$, whereas the terms where $i$ is bound to a ligand~$j$ out of
the remaining $N-1$ ligands is $\exp(-\beta\DGij)\, Z_{-i,-j}/Z_0$. With
these definitions, it follows from Eq.~\eqref{eqn:Z} that
\begin{equation}
1 = \frac{Z_{-i}}{Z} + \sum_j e^{-\beta\DGij}
\frac{Z_{-i,-j}}{Z},\qquad\text{for all~$i$},
\label{eqn:sc_Z}
\end{equation}
where the sum is over the $N-1$ ligands $j$ distinct from $i$.  Let $p_i$ be
the probability that ligand~$i$ is unbound:
\begin{equation}
p_i = Z_{-i} / Z.
\end{equation}
In Eq.~\eqref{eqn:sc_Z}, the factor $Z_{-i,-j}/Z$ is the probability that
both $i$~and~$j$ are unbound.  Noting that $Z_{-i,-j}/Z_{-j}$ is the
conditional probability for $i$ to be unbound given that $j$ is unbound, we
can rewrite $Z_{-i,-j}/Z$ as
\begin{equation}
\frac{Z_{-i,-j}}{Z} = \frac{Z_{-i,-j}}{Z_{-j}} \frac{Z_{-j}}{Z}
\approx  \frac{Z_{-i}}{Z} \frac{Z_{-j}}{Z} = p_i p_j.\label{eqn:sc_approx}
\end{equation}
Above, we have approximated the binding of $i$~and~$j$ as uncorrelated
events.  This approximation is accurate if both $i$~and~$j$ have many
possible binding partners with comparable binding free energies.

Using Eqs.~\eqref{eqn:sc_Z}--\eqref{eqn:sc_approx}, we obtain the following
set of $N$ self-consistent equations for the quantities $\{p_i\}$:
\begin{equation}
1 = p_i + \sum_j e^{-\beta\DGij} p_i p_j,\qquad \text{for all~$i$}.
\end{equation}
When rewritten in the following convenient form, these equations can be
solved, for example, by self-consistent iteration:
\begin{equation}
p_i = \frac{1}{1 + \sum_j e^{-\beta\DGij} p_j},\qquad \text{for all~$i$}.
\label{eqn:sc_pi}
\end{equation}
The probability $p_{ij}$ that ligands $i$~and~$j$ are bound is then given
by
\begin{equation}
p_{ij} = e^{-\beta\DGij} p_i p_j.\label{eqn:sc_pij}
\end{equation}

Eqs.\ \eqref{eqn:sc_pi}~and~\eqref{eqn:sc_pij} are a principal result of
this paper.  It can be shown that these results are a single-particle analog
of the more familiar chemical equilibrium conditions between the
concentrations of $N$ species of particles, with binding constants
proportional to $\exp(-\beta\DGij)$.

Eq.~\eqref{eqn:sc_pi} can be recast as a minimization problem, which leads
to a more efficient method of solution than fixed-point iteration.  A
similar observation has recently been made when solving the equations that
arise in umbrella sampling free energy calculations (WHAM/MBAR), which have
a similar structure~\cite{TanEtAl:2012, ZhuHummer:2012}. Additionally, this
form allows us to prove that Eq.~\eqref{eqn:sc_pi} has exactly one solution,
which satisfies $0 < p_i \leq 1$ for all~$i$ by construction. We define
$f(\{p_i\})$, the function to be minimized, as follows:
\begin{equation}
f(\{p_i\}) = \sum_i [p_i - \ln p_i] + \frac12 \sum_{i,j} p_i k_{ij} p_j,
\end{equation}
where $k_{ij} = \exp(-\beta\DGij)$.  The stationary points of $f$ are given
by the solutions to the following system of equations,
\begin{equation}
0 = \frac{\partial f}{\partial p_i} = 1 - \frac{1}{p_i} + \sum_j k_{ij} p_j,
\end{equation}
which are equivalent to Eq.~\eqref{eqn:sc_pi}.  We now demonstrate that
\emph{all} stationary points of $f$ are local minima, implying there is a
single stationary point that is the global minimum.  To this end, we prove
that the Hessian of $f$ at a stationary point, $\partial^2 f/\partial p_i\,
\partial p_j$, is positive definite, implying that the stationary point is a
local minimum, by showing that the quantity $\sum_{ij} v_i (\partial^2
f/\partial p_i\, \partial p_j) v_j$ is positive for all nonzero
$N$-component vectors $\{v_i\}$.  Indeed,
\begin{align}
\sum_{ij} v_i \frac{\partial^2 f}{\partial p_i\, \partial p_j} v_j
&= \sum_{ij} v_i \left[ \frac{1}{p_i^2}\delta_{ij} + k_{ij} \right] v_j,\\
&= \sum_{ij} \frac{v_i}{p_i} \left[ \delta_{ij} + p_i k_{ij} p_j \right]
\frac{v_j}{p_j}.
\label{eqn:HessFPosDef}
\end{align}
The non-negative quantities in square brackets are the elements of a strictly
diagonally dominant, symmetric matrix with positive diagonal elements, which
is thus positive definite~\cite{Taussky:1949}.  Indeed, the $i^{\text{th}}$
element of its diagonal is equal to $1 + p_i^2 k_{ii} \geq 1$ (we allow
nonzero $k_{ii}$ here, in anticipation of the mean-field self-consistent
theory of Eq.~\eqref{eqn:scmf_pa} below), while the sum of the
off-diagonal terms in the $i^{\text{th}}$ row satisfies
\begin{equation}
\sum_j p_i k_{ij} p_j - p_i^2 k_{ii} \leq \sum_j p_i k_{ij} p_j = 1 - p_i < 1,
\end{equation}
where the equality and the strict inequality both follow from
Eq.~\eqref{eqn:sc_pi} at the stationary point.  Hence, the right-hand
side of Eq.~\eqref{eqn:HessFPosDef} is positive for every nontrivial choice
of $\{v_i\}$, so the Hessian of $f$ is positive definite, as claimed.

Having calculated the probabilities of each bond forming
(Eq.~\eqref{eqn:sc_pij}), we now calculate the resulting attraction free
energy, $F_\att$, given by
\begin{equation}
\beta F_\att = -\ln( Z / Z_0 ).
\end{equation}
This quantity is most easily estimated using thermodynamic integration.  One
possible integration path is to offset all interaction energies by an amount
$\lambda/\beta$, i.e.
\begin{equation}
\beta\DGij \mapsto \beta\DGij + \lambda.\label{eqn:lambda}
\end{equation}
Let $Z(\lambda)$ be the partition function after this replacement. When
$\lambda$ is infinite, $Z(\lambda) = Z_0$, so
\begin{equation}
\beta F_\att
= \int_\infty^0 \dee\lambda\, \frac{\partial[-\ln(Z(\lambda)/Z_0)]}{\partial \lambda}
= -\int_0^\infty \dee\lambda\, \sum_{i<j} p_{ij}(\lambda),
\label{eqn:Fatt_basic}
\end{equation}
where $p_{ij}(\lambda)$ is given by
Eqs.\ \eqref{eqn:sc_pi}~and~\eqref{eqn:sc_pij} after effecting the
substitution of Eq.~\eqref{eqn:lambda}.

Once the free energy is known at one temperature, a more physical
integration path can be used to obtain it at any other temperature.  To
proceed, we denote the dependence of bond energies on $\beta$ explicitly
with the notation $\Delta G_{ij}(\beta)$, and similarly for $p_{ij}(\beta)$
and $F_\att(\beta)$.  We obtain
\begin{equation}
\beta_1 F_\att(\beta_1) - \beta_0 F_\att(\beta_0)
= \int_{\beta_0}^{\beta_1} \dee\beta \sum_{i<j}
\frac{\partial \beta \Delta G_{ij}(\beta)}{\partial \beta} p_{ij}(\beta).
\label{eqn:FattBeta}
\end{equation}
In calculating the integral in Eq.~\eqref{eqn:FattBeta} from
$\beta_0$~to~$\beta_1$, the free energy of binding at all intermediate
temperatures is obtained as well.

Once $F_\att$ is known, the free energy of entropic repulsion,
$F_\rep$, must be added to obtain a full interaction free energy
(see Appendix~\ref{app:entropic} for details).  Additionally, it is usually
convenient to offset all free energies to yield zero total energy when the
colloids are an infinite distance apart, which we do in
Section~\ref{sec:results}.

In the limit of weak binding, when all the factors $\exp(-\beta\DGij)$ are
very small, the above results simplify significantly, as all the
complications of binding competition disappear. Discarding all terms of
order $\exp(-2\beta\DGij)$ and higher, we obtain
\begin{align}
p_i &\approx 1 - \sum_j e^{-\beta\DGij},\\
p_{ij} &\approx \exp(-\beta\DGij),\\
\beta  f_\att &\approx -\sum_{i<j} p_{ij}.
\end{align}
In particular, the last relation can be used to sidestep the thermodynamic
integration in Eq.~\eqref{eqn:Fatt_basic}, and is related to a similar
``Poisson'' approximation used in the
literature~\cite{BiancanielloKimCrocker:2005, LicataTkachenko:2006,
  DreyfusEtAl:2010, RogersCrocker:2011, RogersCrocker:2012}.  However, we
have previously pointed out that under realistic experimental conditions,
this weak-binding approximation is rather poor~\cite{MognettiEtAl:2012}, so
its utility for experiments is limited.

\subsection{Specializing to DNA-coated colloids}
\label{sec:theory:dnaccs}

For a large class of ligands, binding occurs only in a small portion of the
ligand.  In DNACCs in particular, the DNA strands that are used consist of a
long, inert tether capped by a reactive, short sticky end, shown in
Figure~\ref{fig:system}(b).  The thermodynamics of association of DNA
oligomers is well understood~\cite{SantaLuciaJr.Hicks:2004}, so for a pair
of strands, $i$~and~$j$, we can predict from their sequence the value of the
hybridization free energy of their sticky ends in solution, $\DGij^0$.
Separating out this contribution to $\DGij$, we have
\begin{equation}
\DGij = \DGij^0 + \DGij^\cnf,
\label{eq:separation}
\end{equation}
where $\DGij^\cnf$ is the free energy penalty that arises from the reduced
configurational space available to two bound ligands with respect to two
free ligands\cite{DreyfusEtAl:2009, LeunissenFrenkel:2011}.  This quantity
depends only on the polymer statistics of the tethers and the way in which
they are attached to the colloids (e.g., tethered on the surface
vs. mobile), but not on the sticky ends, and in many cases is a completely
entropic effect.  In Appendix~\ref{app:entropic}, we list explicit
expressions for $\DGij^\cnf$ applicable to short, rigid dsDNA tethers
grafted on plates. We also show how to use common polymer statistics
techniques to estimate $\DGij^\cnf$ to arbitrary accuracy for more
complicated tethers, such as long, flexible ssDNAs modeled as ideal
chains~\cite{BiancanielloKimCrocker:2005, RogersCrocker:2011}.

\subsection{Mean field theory for DNA-coated plate interactions}
\label{sec:theory:meanfield}

In some regimes, notably micron-sized colloids with short DNA tethers, the
range of the binding interactions is small compared to the radii of
curvature of all the colloidal surfaces involved.  In such cases, we can use
the Derjaguin approximation~\cite{Hunter:2000} to calculate particle
interaction energies, knowing only the interaction energy density of a pair
of uniformly coated parallel plates as a function of plate separation.  For
this translationally invariant geometry, a substantial simplification is
possible if the ligands are treated at a mean field level rather than
individually.

The simplified system consists of two parallel plates separated by a
distance~$h$.  We consider $M$ different types of ligands, labeled by
letters $a$~and~$b$ below.  Two ligand types can differ in their sticky end
sequences, tether types or in the plate on which they are grafted.
Corresponding to these $M$ types, there are $M (M+1)/2$ potentially
different solution hybridization free energies, denoted by $\Delta
G^0_{ab}$.  Let $\sigma_a$ be the grafting density of $a$-type tethers,
which has units of tethers per unit area.  Since we approximated the system
as uniform, all $a$-type tethers have the same probability of being unbound,
denoted by~$p_a$.  By approximating the sum in Eq.~\eqref{eqn:sc_pi} by a
continuous integral, we obtain a mean-field, self-consistent set of
equations
\begin{equation}
p_a = \frac{1}{1 + \sum_b K_{ab} \sigma_b p_b},\label{eqn:scmf_pa}
\end{equation}
where
\begin{equation}
K_{ab} = e^{-{\beta \DG^0_{ab}}}\,
\int\dee^2\vr_b\,\exp[-\beta\DG_{ab}^\cnf].\label{eqn:scmf_Kab}
\end{equation}
In the last equation, $\DG_{ab}^\cnf$ is the configurational cost for a
fixed $a$-type ligand to bind a $b$-type tether grafted at~$\vr_b$.  The
factor $K_{ab}$ has units of area, and can be interpreted as an average
Boltzmann factor for tether binding times the area on plate~$B$ within reach
of a fixed $a$-type tether.  A relation analogous to Eq.~\eqref{eqn:sc_pij}
yields the density of bonds between $a$-type and $b$-type tethers,
$\sigma_{ab}$:
\begin{equation}
\sigma_{ab} = \sigma_a p_a K_{ab} p_b \sigma_b.\label{eqn:scmf_pab}
\end{equation}
Similarly, using a relation analogous to Eq.~\eqref{eqn:Fatt_basic}, we can
use thermodynamic integration to estimate the free energy of attraction per
unit area, $f_\att(h)$, at fixed plate separation:
\begin{equation}
\beta f_\att(h) = -\int_0^\infty\dee\lambda\, \sum_{a\leq b}
\sigma_{ab}(\lambda),
\label{eqn:scmf_f}
\end{equation}
where $\lambda$ is the integration parameter defined in
Eq.~\eqref{eqn:lambda}.

Eqs. \eqref{eqn:scmf_pa}--\eqref{eqn:scmf_f} are a second principal result
of this paper.  They constitute a self-consistent mean field theory for the
DNA-mediated attraction of uniformly coated parallel plates.  In
Appendix~\ref{app:saddlepoint}, we show that, when specialized to only one
or two kinds of bonds, these equations reduce to the earlier self-consistent
mean field theories of Refs.~\onlinecite{MognettiLeunissenFrenkel:2012,
  Angioletti-UbertiMognettiFrenkel:2012}.  In these particular cases, we
also show how Eqs. \eqref{eqn:scmf_pa}--\eqref{eqn:scmf_f} may be derived
from a saddle-point approximation of a mean-field partition function.

For the case of short, rigid dsDNA tethers, the factors $K_{ab}$ have
particularly simple expressions, derived in Appendix~\ref{app:lce}.  In
particular, let tethers of type $a$~and~$b$ have lengths $L_a$~and~$L_b$.
Without loss of generality, let $L_a < L_b$.  When $a$~and~$b$ are on the
same plate,
\begin{subequations}
\begin{equation}
K_{ab} = \frac{e^{-\beta\DG^0_{ab}}}{\rho_0} \times
\begin{cases}
 1/L_b,&L_b < h;\\
 1/h,&\text{otherwise}.
\end{cases}
\end{equation}
whereas when $a$~and~$b$ are on opposite plates,
\begin{equation}
K_{ab} = \frac{e^{-\beta\DG^0_{ab}}}{\rho_0} \times
\begin{cases}
 0,&L_a + L_b < h;\\
 \frac{L_a + L_b - h}{L_a L_b},&L_b < h < L_a + L_b;\\
 1/h,&\text{otherwise}.
\end{cases}
\end{equation}
\label{eqn:simpleKab}
\end{subequations}

When the tethers are modeled as freely-jointed chains of segments of
length~$\ell$, a choice that has been suggested for modeling ssDNA
tethers~\cite{RogersCrocker:2011}, the factors $K_{ab}$ can be approximated
with the following expressions, derived in
Appendix~\ref{app:entropic:polymers}.  Let $N_a$~and~$N_b$ be the number of
segments in tethers of type $a$~and~$b$.  When $a$~and~$b$ are on the
same plate,
\begin{subequations}
\begin{multline}
K_{ab} \approx \frac{e^{-\beta\DG^0_{ab}}}{\rho_0}\\
\times\sqrt{\frac{3}{\pi (N_a + N_b) \ell^2}}
\erf\Bigl(\sqrt{\frac{3 h^2}{(N_a + N_b) \ell^2}}\Bigr)\\
\Bigg/ \erf\Bigl(\sqrt{\frac{3 h^2}{2 N_a \ell^2}}\Bigr)
\erf\Bigl(\sqrt{\frac{3 h^2}{2 N_b \ell^2}}\Bigr),
\end{multline}
whereas when $a$~and~$b$ are on opposite plates,
\begin{multline}
K_{ab} \approx \frac{e^{-\beta\DG^0_{ab}}}{\rho_0}
\exp\Bigl[-\frac{3 h^2}{4 \ell^2 (N_a + N_b)}\Bigr]\\
\times\sqrt{\frac{12}{\pi (N_a + N_b) \ell^2}}
\erf\Bigl(\sqrt{\frac{3 h^2}{4 (N_a + N_b) \ell^2}}\Bigr)\\
\Bigg/ \erf\Bigl(\sqrt{\frac{3 h^2}{2 N_a \ell^2}}\Bigr)
\erf\Bigl(\sqrt{\frac{3 h^2}{2 N_b \ell^2}}\Bigr).
\end{multline}
\label{eqn:simpleKabFJC}
\end{subequations}
In both of these equations, the error function, $\erf(x)$, is defined as
\begin{equation}
\erf(x) = \frac{2}{\pi} \int_0^x \dee t\, e^{-t^2}.
\label{eqn:erf_defn}
\end{equation}

As before, when all the factors $K_{ab} \sigma_b$ and $\sigma_a K_{ab}$ are
very small, one can implement a weak-binding approximation to sidestep the
thermodynamic integration in Eq.~\eqref{eqn:Fatt_basic}, which yields
\begin{align}
p_a &\approx 1 - \sum_b K_{ab} \sigma_b,\\
\sigma_{ab} &\approx \sigma_a K_{ab} \sigma_b,\\
\beta f_\att &\approx -\sum_{a\leq b} \sigma_{ab}.
\end{align}
However, as pointed out previously, the utility of this approximation under
realistic experimental conditions is limited.

Frequently, DNACCs are covered uniformly at a fixed density~$\sigma$ with a
single type of strand, and only the sticky ends vary from one colloid to the
next.  In this case, we can obtain an explicit mean-field expression for
$\beta f_\att$ between a plate grafted with $a$-type strands and one grafted
with $b$-type strands.  Defining $J$ as the dimensionless factor $2 K_{ab}
\sigma$, we obtain
\begin{equation}
\sigma_{ab} = \frac{\sigma}{J} \left[ (1 + J) - \sqrt{1 + 2J} \right],
\end{equation}
so,
\begin{multline}
\beta f_\att = \frac{\sigma}{J} \biggl[ -(1 + J) + \sqrt{1 + 2J}\\
+ 2 J \ln\Bigl\{\frac12 (1 + \sqrt{1 + 2J})\Bigr\} \biggr].
\end{multline}
This expression makes explicit the breakdown of the weak-binding
approximation (in this case, $\beta f_\att \approx -\sigma_{ab}$) when $J
\gtrsim 1$.  Unfortunately, even for the simple case of rod-like DNA
strands, the above expression does not yield a simple analytic result for
the pair potential between to DNA-coated spheres via the Derjaguin
approximation.  Since this difficulty only increases when we allow for
uneven coverages and multiple potential interactions, we believe that our
numerical thermodynamic integration procedure is justified.


\section{Results and discussion}
\label{sec:results}

In this section, we compare the results of detailed Monte Carlo simulations
(MC) with those derived from our theory, described above.  We refer to the
results of the theory where each strand is modeled individually (Eqs.
\eqref{eqn:sc_pi}, \eqref{eqn:sc_pij}~and~\eqref{eqn:Fatt_basic}) as the SCT
results, and those of the form where the strands are modeled at a mean field
level (Eqs. \eqref{eqn:scmf_pa}--\eqref{eqn:scmf_f}) as the SCTMF
results\footnote{Self-contained programs to calculate the SCT and SCTMF
  predictions for all of these examples are included in the accompanying
  code, in the \texttt{examples} folder.}.

\subsection{Colloids coated with short dsDNA strands}
\label{sec:results:dsDNA}

The simplest possible DNACC is a flat plate that is uniformly covered by a
single type of DNA strand.  Interaction energies obtained from plate
geometries can be used to approximate interaction energies in more complex
geometries by use of the Derjaguin approximation.  Although, our theory can
be used in arbitrary geometries, in this section we focus on plates for
clarity.

\begin{figure}
\begin{center}
\includegraphics{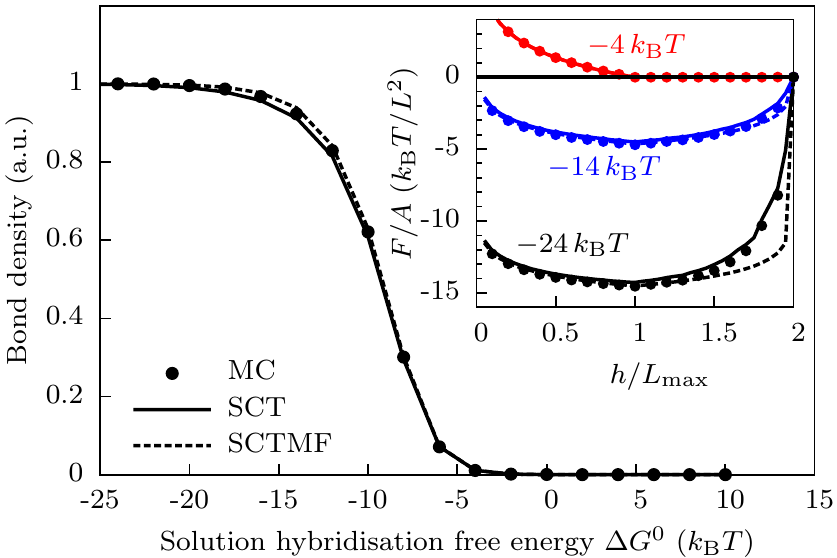}
\end{center}
\caption{\label{fig:a-ap}Density of bonds formed between two
  flat plates coated uniformly with short, rigid dsDNA tethers of length $L
  = 20\,$nm, at plate separation $h = L$, as a function of the sticky end
  solution hybridization free energy, $\DG^0$.  Inset: potential of mean
  force per unit area as a function of plate separation, for three different
values of $\DG^0$.}
\end{figure}

The interaction free energy per unit area of two such DNACCs has been
calculated previously from Monte Carlo
simulations~\cite{LeunissenFrenkel:2011, MognettiLeunissenFrenkel:2012}.  In
this case, the tethers were modeled as short, stiff rods of length $L =
20\,$nm, grafted at a density of $1/(20\,\text{nm})^2$.
Figure~\ref{fig:a-ap} shows the fraction of bonds formed between the two
plates when separated by a distance $h = L$, as a function of the sticky end
solution hybridization free energy, $\DG^0$.  The results are calculated
both from Monte Carlo simulations and from our theory, using an identical
realization of the tether grafting positions on a periodic
$200\,\text{nm}\times 200\,\text{nm}$ plate.  The figure inset shows the
free energy per unit area of these plates, as a function of plate
separation, for three different values of~$\DG^0$.

For the grafting density used in Figure~\ref{fig:a-ap}, each tether
typically has around $9$~possible binding partners, so the mean-field
approximation is excellent.  Small discrepancies arise only at low
temperature and large plate separation, where the number of possible binding
partners is low and Poisson fluctuations in local coverage density are
relevant.  Such fluctuations are naturally accounted for in the explicit
self-consistent theory, whose predictions are thus nearly indistinguishable
from MC simulations in all cases.

\begin{figure}
\begin{center}
\includegraphics{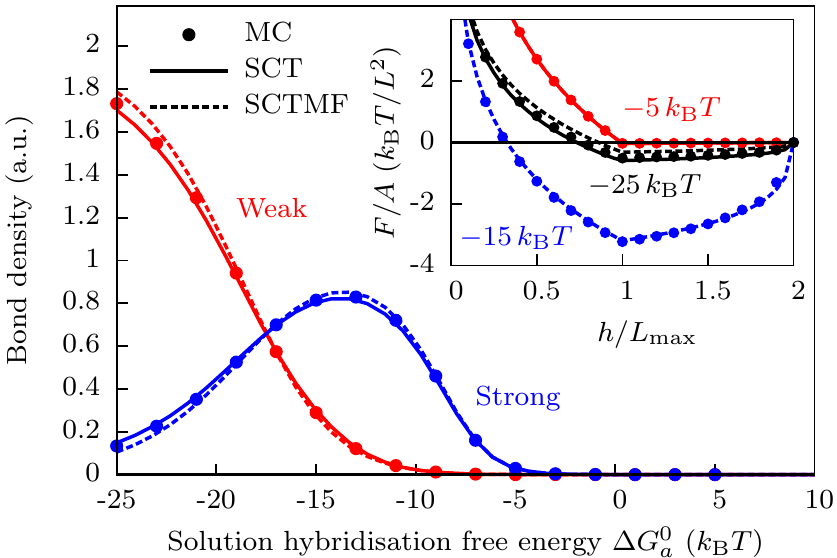}
\end{center}
\caption{\label{fig:ab-apbp}Density of strong and weak bonds formed between
  plates with competing-linkages, at plate separation $20\,$nm.  Strong
  $a$-$a'$ inter-plate bonds (bridges) have sticky end hybridization free
  energy $\Delta G^0_{a}$, and weak $a$-$b$ and $a'$-$b'$ intra-plate bonds
  (loops) have sticky end hybridization free energy $\Delta G^0_{a} +
  5\,k_{\text{B}}T$.  See text for details.  Inset: potential of mean force
  per unit area as a function of plate separation, for $\DG^0_{a} = -5\,\kT$
  (upper), $-15\,\kT$ (bottom) and $-25\,\kT$ (middle).  When the sticky end
  binding is strongest, the plates barely interact.}
\end{figure}

Figure~\ref{fig:ab-apbp} shows the analogous results for a pair of DNACCs
coated with two kinds of tethers which has been described
previously~\cite{Angioletti-UbertiMognettiFrenkel:2012}.  We label the
tethers on one plate by $a$~and~$b$, and those on the other plate by
$a'$~and~$b'$.  Each tether type is grafted at a density of $1 /
(20\,\text{nm})^2$.  The sticky ends of $a$ and $a'$ hybridize strongly,
with free energy $\Delta G^0_{a}$, forming bridges between the two plates.
Conversely, the sticky ends of $a$~and~$b$, as well as those of
$a'$~and~$b'$, hybridize weakly, with free energy $\Delta G^0_{a} + 5\,\kT$,
forming loops within each plate.  A competition thus arises between strong
bridges and weak loops, which has been shown to result in interaction
potentials with an unusual temperature dependence that should significantly
enhance the ability of these DNACCs to form binary
crystals~\cite{Angioletti-UbertiMognettiFrenkel:2012}.  All the features of
these interactions are reproduced with quantitative accuracy by our
self-consistent theory.

The key advantage of the present theory is its generality and comparative
simplicity.  With little additional effort, we can consider systems with
many different kinds of grafted DNAs, each with different entropic costs for
binding, and with different and potentially nonuniform coating densities.
In the previous two cases, the resulting SCTMF theory is equivalent to
earlier self-consistent mean-field theories that were less general and that
were derived more heuristically (Appendix~\ref{app:saddlepoint}). In our
third and fourth examples, we go beyond what has been done previously to
illustrate the generality of our approach.

\begin{figure}
\begin{center}
\includegraphics{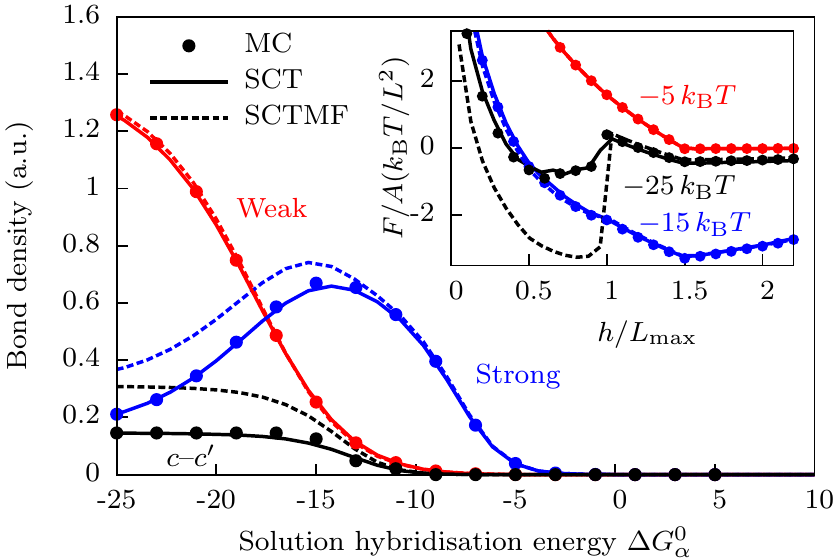}
\end{center}
\caption{\label{fig:abc}Density of strong, weak and $c$--$c'$ bonds formed
  in the third system, at plate separation $20\,$nm (see text for details).
  Inset: potential of mean force per unit area as a function of plate
  separation, for three values of $\DG^0_{a}$.}
\end{figure}

For our third example, we generalize the setup of Figure~\ref{fig:ab-apbp}
to include a third type of tether on each plate, respectively $c$~and~$c'$.
The sticky ends of $c$~and~$c'$ hybridize only with each other and with the
same strength as the $a$~and~$a'$ sticky ends.  Furthermore, we set the
lengths of the $a$, $b$, $a'$~and~$b'$ tethers to $30\,$nm and those of the
$c$~and~$c'$ tethers to $10\,$nm.  Finally, we set the grafting density of
the $c$~and~$c'$ tethers to $1 / (30\,\text{nm})^2$.  Thus, our third system
features three different types of rods, of different lengths, with different
coverages and with a nontrivial interaction matrix.  Figure~\ref{fig:abc}
summarizes the behavior of this system.  Our choices result in the
interaction energy per unit area having up to two different minima,
depending on the specific temperature (binding strength).  The example also
illustrates one way in which the mean-field approach fails, and how the full
theory behaves for such cases.  The large discrepancies in the number of
$c$--$c'$ bonds formed (black solid and dashed curves in
Figure~\ref{fig:abc}), and the resulting discrepancies in the potentials of
mean force, result from the low coverage density of these tethers.  The
typical separation between grafting points that result from Poisson
statistics is far larger than the length of the tethers, but these
inhomogeneities are not captured by a mean field approach.  Importantly,
since the full theory retains detailed information on the grafting points of
$c$~and~$c'$ tethers, the MC results are again recovered to quantitative
accuracy.

\begin{figure}
\begin{center}
\includegraphics{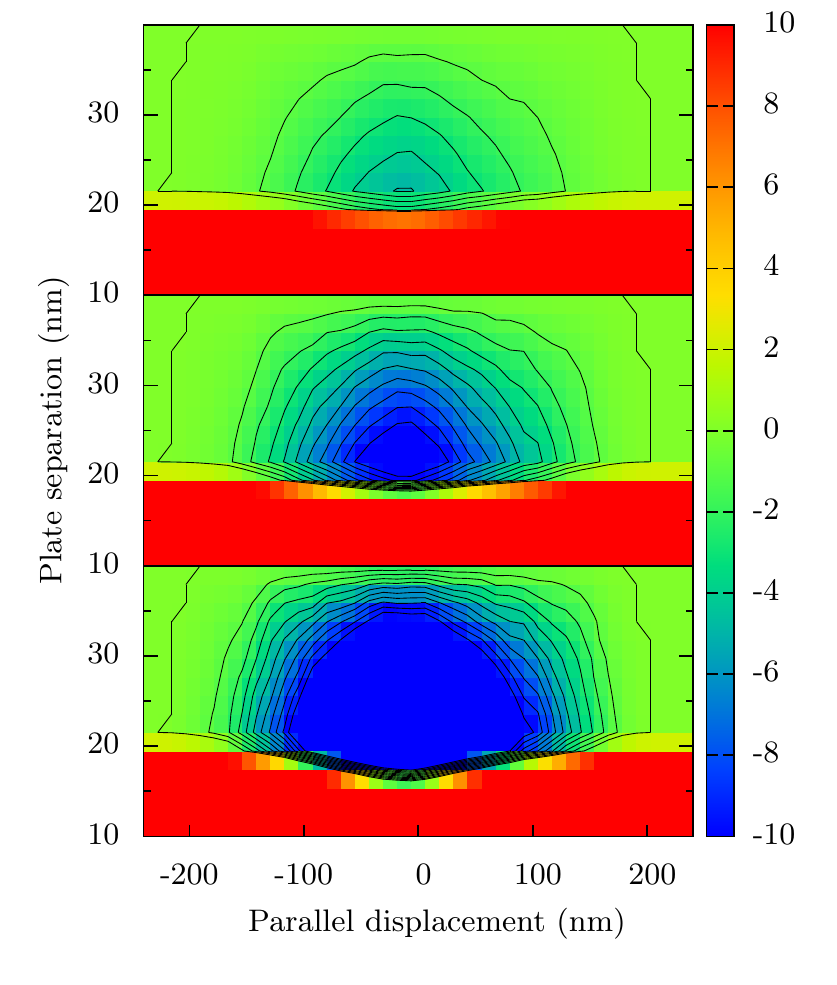}
\end{center}
\caption{\label{fig:patches}Potential of mean force for two parallel plates
  coated with circular patches (radius $100\,$nm) of short dsDNA strands
  with complementary sticky ends.  The coating density is $1/(20\,$nm$)^2$.
  The sticky ends have solution hybridization free energy of $-6\,\kT$
  (top), $-7\,\kT$ (middle) and $-8\,\kT$ (bottom).  Contours are drawn with
  $1\,\kT$ spacing.}
\end{figure}

Our fourth and final example in this section is a system of two parallel
plates, each with a $100\,$nm-radius circular patch of dsDNA strands,
distributed uniformly within the patch at a coverage density
$1/(20\,$nm$)^2$.  As above, the strands are modeled as $20\,$nm rigid rods.
The only hybridization allowed is between strands on opposite plates, and we
consider solution hybridization energies of $-8\,\kT$, $-7\,\kT$ and
$-6\,\kT$.  Figure~\ref{fig:patches} shows the potentials of mean force of
these plates as a function of plate separation and lateral displacement
(zero lateral displacement means that the centers of the circular patches
line up vertically).  Such detailed potentials could be used to inform
parameter choices for coarse-grained models of patchy
particles~\cite{KernFrenkel:2003} when the patches are made from
patterned DNA coatings.  For nanoscopic colloids, we suggest that such
patches could be realized by building the particles through DNA
origami\cite{Rothemund:2006}.  However, to our knowledge, such precise
experimental control of coating patterns for microscopic colloids remains
elusive, although recent experiments with triblock Janus particles are very
encouraging\cite{ChenBaeGranick:2011}.

\subsection{Colloids coated with long ssDNA tethers}
\label{sec:results:ssDNA}

To describe more flexible ssDNA tethers, we use the representation recently
employed by Rogers and Crocker~\cite{RogersCrocker:2011} to describe
experimental measurements of DNACC interaction potentials.  They model ssDNA
tethers of contour length $40\,$nm as 8-segment freely jointed chains of
Kuhn length $5\,$nm.  The sticky ends that cap them are modeled as
point-like, so that when two tethers bind, they form a 16-segment freely
jointed chain.  Appendix~\ref{app:entropic:polymers} describes how to
calculate $\DG^\cnf_{ij}$ for two such tethers $i$~and~$j$.

In Rogers and Crocker's experiment, there are two complementary types of
sticky ends, $A$~and~$B$, whose sequences are predicted\footnote{An
  ambiguity arises from including or excluding the bases of the tether
  immediately adjacent to the sticky end, since they can form stacking
  interactions with the dsDNA formed by two bound sticky ends.  If included,
  DINAmelt predicts $\Delta H^0 = -56.7\,$kcal/mol and $\Delta S^0 =
  -166.0\,$cal/(mol$\cdot$K).  Since the scope of this paper is to evaluate
  the accuracy of the SCT treatment with respect to full MC simulations, we
  sidestep any attempt to resolve this ambiguity.} by
DINAmelt~\cite{MarkhamZuker:2005} to have binding enthalpy $\Delta H^0 =
-47.8\,$kcal/mol and binding entropy $\Delta S^0 =
-139.6\,$cal/(mol$\cdot$K).  At temperature $T$, the $A$-$B$ binding free
energy in solution is taken to be $\Delta G^0 = \Delta H^0 - T \Delta S^0$.
Tethers with these sticky ends are coated on three types of spherical
colloids, all of diameter $1.1\,\mu$m, $A$-type colloids are coated with
$4800\pm480$ $A$-type tethers, $B$-type colloids are coated with
$4200\pm420$ $B$-type tethers, and $AB$-type colloids are coated with
$2400\pm240$ $A$-type and $3500\pm350$ $B$-type tethers.  In all the
following simulations, we have used the central value of all ranges.

\begin{figure}
\begin{center}
\includegraphics{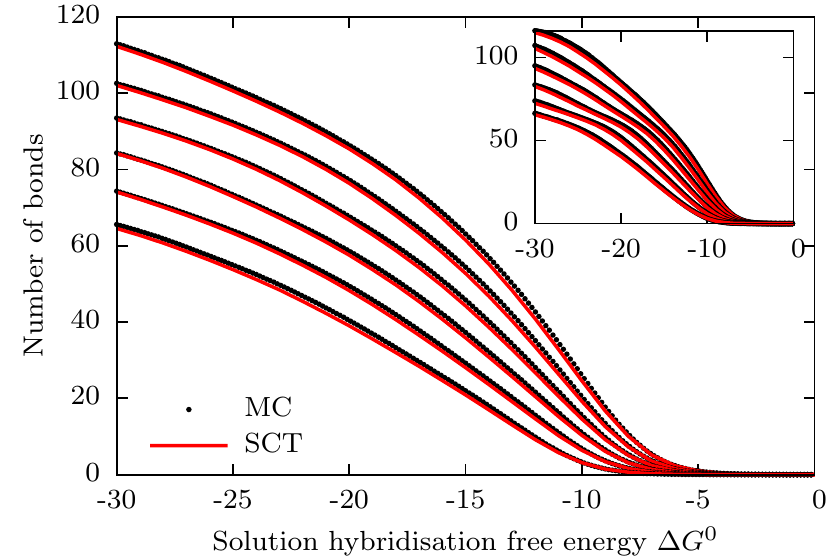}
\end{center}
\caption{\label{fig:AvgNumBondsAB}Calculated number of bonds formed between
  the $A$- and $B$-type colloids described in
  Ref.~\onlinecite{RogersCrocker:2011}, when separated by $5\,$nm (top curve) to
  $30\,$nm (bottom curve) in steps of $5\,$nm, averaged over $14$
  realizations of tether grafting points on the colloids.  Inset: Same
  before averaging, i.e., for an individual realization of tether grafting
  points.}
\end{figure}

We first address the interaction between $A$~and~$B$ colloids.
The coating densities used by Rogers and Crocker are sufficiently low that
local grafting density fluctuations are relevant.  We have thus prepared
$14$~independent configurations of uniformly coated $A$~and~$B$ colloids.
Figure~\ref{fig:AvgNumBondsAB} shows the number of bonds formed
between two such colloids when separated by $5\,$nm to $30\,$nm versus
sticky end binding strength $\Delta G^0$, as calculated by explicit Monte
Carlo simulation and the explicit SCT theory, averaged over the $14$
independent coating realizations.  The agreement is nearly quantitative,
both when averaged over grafting realizations and independently for each
such realization (inset).

\begin{figure}
\begin{center}
\includegraphics{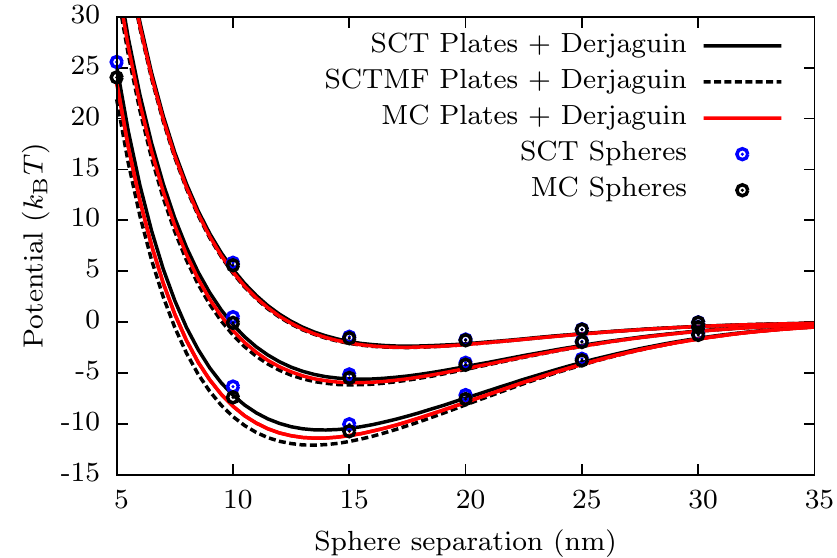}
\end{center}
\caption{\label{fig:ABSpherePotentials}Potential of mean force between $A$
  and $B$ colloids of Ref.~\onlinecite{RogersCrocker:2011} at temperatures of
  $36.0^\circ$C, $33.0^\circ$C and $30.5^\circ$C (upper, middle and lower
  curves).  See text for details of the various methods.}
\end{figure}

\begin{figure}
\begin{center}
\includegraphics{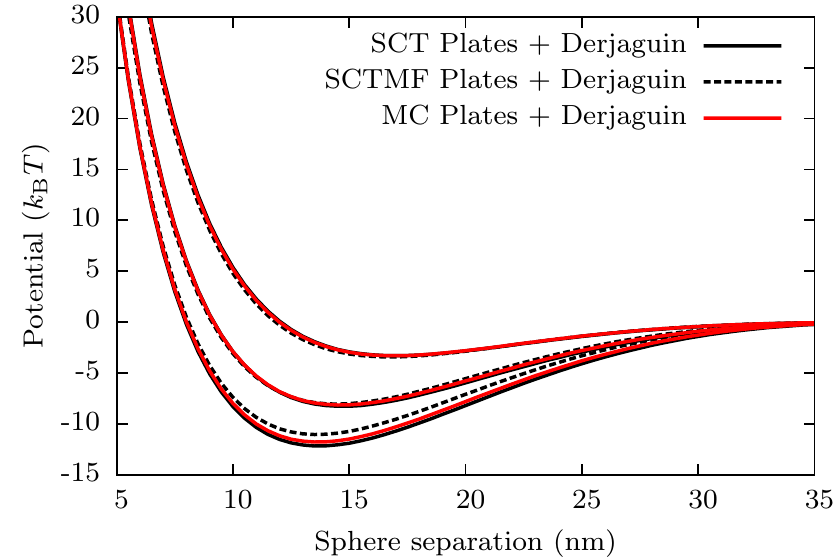}
\end{center}
\caption{\label{fig:ABABSpherePotentials}Potential of mean force between two
  $AB$ colloids of Ref.~\onlinecite{RogersCrocker:2011} at temperatures of
  $32.0^\circ$C, $27.0^\circ$C and $24.0^\circ$C (upper, middle and lower
  curves).  See text for details.}
\end{figure}

By using Eq.~\eqref{eqn:Fatt_basic}, we can integrate the curves in
Figure~\ref{fig:AvgNumBondsAB} to obtain the mean attraction between the
$A$~and~$B$ colloids.  Further accounting for the entropic repulsion caused
by confining the tethers between the two colloids results in a potential of
mean force for this system, shown as dots in
Figure~\ref{fig:ABSpherePotentials}.  In this Figure, we also show the
result of calculating the colloid interaction potential by calculating the
interaction energy per unit area between two similarly coated plates, and
then applying the Derjaguin approximation.  The interaction free energy
between plates is calculated in three different ways: by explicit MC
simulations, by using SCT, and by using its mean-field approximation, SCTMF.
Our results here help establish the Derjaguin approximation as valid for
ssDNA tethers modeled as freely-jointed chains.  For completeness, we show
similar results for the AB--AB system explored by Rogers and Crocker in
Figure~\ref{fig:ABABSpherePotentials}.

It should be noted that the simple freely-jointed chain model of ssDNA used
here and in Ref.~\onlinecite{RogersCrocker:2011} may not adequately capture the
complications of the experimental system~\cite{MognettiEtAl:2012,
  RogersCrocker:2012}.  Our aim in this paper is limited to
establishing the SCT as an accurate and fast alternative to detailed MC
calculations, regardless of the underlying model used to calculate the
configurational entropic factors associated with individual tethers binding.


\section{Summary of results}
\label{sec:summary}

For ease of reference, in this section we summarize the main results of the
paper.  These results are implemented in the Python package \verb+DNACC+,
available at \verb+http://github.com/patvarilly/DNACC+.

\subsection{Self-consistent theory (SCT)}

Our central result describes the interaction details of a DNACC system where
each strand~$i$ is modeled explicitly.  The algorithm is:
\begin{enumerate}
\item Using a suitable microscopic model, calculate the binding free energy
  $\DGij$ between every pair of strands $i$~and~$j$.  When $i$~equals~$j$,
  or when $i$~and~$j$ cannot bind, set $\DGij$ to infinity.  This binding
  free energy is the sum of two contributions: the solution hybridization
  free energy of the sticky ends, $\DG^0_{ij}$, and the entropic cost of
  binding of the two tethers that hold the sticky ends, $\DGij^\cnf$. We
  provide expressions for $\DGij^\cnf$ for rod-like and polymeric tethers
  in Appendix~\ref{app:entropic}.

\item Solve the self-consistent set of equations for each of the quantities
  $p_i$, i.e., the probability $p_i$ that strand~$i$ is unbound:
\begin{equation}
p_i = \frac{1}{1 + \sum_j e^{-\beta\DGij} p_j},\qquad \text{for all~$i$}.
\end{equation}
It is straightforward to solve these equations using fixed-point iteration,
i.e., making an initial guess for $\{p_i\}$, then repeatedly inserting the
values of $\{p_i\}$ into the right-hand side of the above equations to
obtain an improved estimate for $\{p_i\}$.  A solution can also be obtained
through minimization of a function with no other local minima or saddle
points, as detailed in Section~\ref{sec:theory:statmech}.

\item Calculate $p_{ij}$, the probability $p_{ij}$ that strands $i$~and~$j$
  are bound:
\begin{equation}
p_{ij} = e^{-\beta\DGij} p_i p_j.
\end{equation}
\end{enumerate}

The interaction free energy of a DNACC system can then be calculated through
thermodynamic integration of the binding free energies $\DGij$, as follows:
\begin{enumerate}
\item Calculate the the free energy of repulsion, $F_\rep$, equal to the
  free energy of the system when none of the strands can bind, i.e., when
  all the $\DGij$ are infinite.  This is simply the sum of the free energies
  of repulsion of the individual strands.  We assume this quantity can be
  calculated analytically or otherwise, as described in
  Appendix~\ref{app:entropic}.

\item Define a thermodynamic integration variable, $\lambda$, and denote by
  $p_{ij}(\lambda)$ the probability that strands $i$~and~$j$ are bound when
  the all the binding free energies are respectively set to $\DGij +
  \lambda\,\kT$, calculated as above.

\item The free energy of the system is given by the following equation
\begin{equation}
F = F_\rep - \kT\,\int_0^\infty\dee\lambda\,\sum_{i<j} p_{ij}(\lambda).
\end{equation}
\end{enumerate}

\subsection{Self-consistent mean-field theory (SCMFT)}

When colloids are coated uniformly with a small number of different DNA
species at high density, it is useful to model the colloid interaction in
mean field.  Assuming that the colloidal surfaces have radii of curvature
significantly larger than the range of interaction of two strands, we may
first calculate the interaction free energy of two parallel plates as a
function of plate separation, $h$, then apply the Derjaguin approximation.

The details of the interaction between two parallel DNA-coated plates
separated by a distance~$h$ are estimated as follows:
\begin{enumerate}
\item For each pair of DNA species, $a$~and~$b$, calculate a binding
  factor $K_{ab}$ as follows:
\begin{equation}
K_{ab} = e^{-{\beta \DG^0_{ab}}}\,
\int\dee^2\vr_b\,\exp[-\beta\DG_{ab}^\cnf],
\end{equation}
where $\DG_{ab}^\cnf$ is the configurational cost for a fixed $a$-type
ligand to bind a $b$-type tether grafted at~$\vr_b$.  If a DNA species is
present in both parallel plates, treat the DNA on each plate as a separate
species.  Simple analytic expressions for $K_{ab}$ for rod-like and
polymeric tethers are given in
Eqs. \eqref{eqn:simpleKab}~and~\eqref{eqn:simpleKabFJC}.

\item Solve the self-consistent set of equations for each of the quantities
  $p_a$, i.e., the probability that an $a$-type strand is unbound:
\begin{equation}
p_a = \frac{1}{1 + \sum_b K_{ab} \sigma_b p_b},
\end{equation}
where $\sigma_b$ is the number of $b$-type strands per unit area.

\item Calculate the number of $a$-~and~$b$-type bonds per unit are,
  $\sigma_{ab}$, as follows:
\begin{equation}
\sigma_{ab} = \sigma_a p_a K_{ab} p_b \sigma_b.
\end{equation}
\end{enumerate}

The free energy per unit area between two DNA-coated parallel plates
separated by a distance $h$ is estimated as follows:
\begin{enumerate}
\item Calculate the free energy density of repulsion, $f_\rep(h)$, equal to
  the free energies of repulsion of the individual strand types, weighted by
  their coverage density (Eq.~\eqref{eqn:frepMFT}).  Expressions for this
  quantity in the case of rod-like and polymeric ligands are given in
  Eqs. \eqref{eqn:frepMFT}, \eqref{eqn:FrepiRods}~and~\eqref{eqn:FrepiFJC}.

\item Define a thermodynamic integration variable, $\lambda$, and denote by
  $\sigma_{ab}(\lambda)$ the number of $a$--$b$ bonds per unit area when the
  all the solution binding free energies are respectively set to $\DG^0_{ab}
  + \lambda\,\kT$, calculated as above.

\item The free energy density of the system is given by the following
  equation
\begin{equation}
f(h) = f_\rep(h) - \kT\,\int_0^\infty\dee\lambda\,\sum_{a \leq b}
\sigma_{ab}(\lambda).
\end{equation}
\end{enumerate}

Finally, to estimate the free energy of interaction between arbitrary
colloids, use the Derjaguin approximation~\cite{Hunter:2000}.  In the case
of two spheres of radius $R_1$~and~$R_2$, respectively, with a distance of
closest approach~$h$, this approximation yields:
\begin{equation}
F(h) \approx \frac{2\pi}{R_1^{-1} + R_2^{-1}} \int_h^\infty\dee h'\, f(h').
\end{equation}
In using the Derjaguin approximation, it is important to shift all free
energies appropriately to ensure that $f(h)$ is zero at infinite~$h$.


\section{Conclusions and Outlook}
\label{sec:conclusions}

The theory for DNACC interactions that we have presented, both in its
explicit form (Eqs. \eqref{eqn:sc_pi},
\eqref{eqn:sc_pij}~and~\eqref{eqn:Fatt_basic}) and its mean-field
approximation (Eqs. \eqref{eqn:scmf_pa}--\eqref{eqn:scmf_f}) is capable
of describing many different DNACC setups, varying in grafting density,
sticky end interactions and tether details. The theory is robust and
correctly incorporates the statistical mechanics of valence-limited binding,
which plays a key role in the behavior of DNACCs, and it does so at
much lower cost than full MC simulations.  Extending the theory to treat
general ligands within our approximation is straightforward, as exemplified
in Appendix~\ref{app:entropic} for freely-jointed-chains.

When comparing our full theory to MC data in the context of DNACCs, we have
found quantitative agreement.  The mean-field approximation, which
substantially generalizes earlier approaches, is in quantitative to
semi-quantitative agreement with the exact results.  Significant
discrepancies arise only when the grafting density fluctuations that are
neglected in the mean-field approach become relevant.

Our theory accounts for all the complications arising from valence
limitation and combinatorial entropy.  Nevertheless, the binding free
energy~$\DGij$ of a pair of ligands $i$~and~$j$ is taken as an external
input, and the agreement of the theory's predictions with experimental
results will depend significantly on the quality of the estimates for the
values of~$\DGij$.  Simple models for generating these estimates, such as
those used in this paper, can inform experiments, though apparently, they
are not yet sufficiently accurate for quantitative agreement with
experiments, as we have discussed in Ref.~\onlinecite{MognettiEtAl:2012}.

We expect our theory to be a significant aid in the search for viable
approaches to design crystal structures \emph{a priori}.  Presently,
existing free energy methods can establish the stable crystal structure of a
system given its interaction potential~\cite{FrenkelSmit:2001}.  Conversely,
robust methods are being developed for solving the inverse problem,
determining the interaction potentials that might yield a particular target
structure~\cite{Torquato:2009}.  Designing colloids that realize such
interaction potentials experimentally is another problem.  In this regard,
DNACCs, with their highly tunable interactions and vast parameter space
(e.g., particle size, ligand structure and flexibility, coating density,
nucleotide sequence length and hybridization strength), are a promising tool
to build colloids with programmable interactions.  For this reason, our
theory can complement existing tools in crystal structure design by quickly
suggesting experimentally realizable constructs that result in particular
colloid--colloid interaction potentials.  Whereas for simple potentials one
can envision a manual search of parameter space guided by intuition and
experience, it might also be possible to recast the problem in terms of a
global minimization algorithm that optimizes the DNACC structures to obtain
interaction potentials that are close to some target.  It is precisely to
help others easily use our tools in these and other ways that we have
written a comprehensive and well documented series of Python modules that
implement our theory, available under the GNU license at
\url{http://github.com/patvarilly/DNACC}.

While we have developed our theory and the accompanying programs with an eye
towards applying them to DNA-coated colloids, other interesting applications
are possible.  In particular, our theory can help model any system where
attraction is dominated by linker-receptor-type bonding.  These include
patchy particles, where the patches are realized by a nonuniform ligand
coat.  More generally, many biological systems could be treated with our
theory, for example, interactions between cells (or between cells and a
substrate) that are mediated by membrane proteins.  For these more general
cases, care should be taken to include the effects of surface elasticity and
how the entropic costs of ligand binding are, in turn, modified.

\begin{acknowledgements}
We thank M.~Leunissen and F.~Martinez-Veracoechea for useful discussions and
a critical reading of the manuscript.  We also thank G.~Hummer for pointing
out the analogy between Eqs. \eqref{eqn:sc_pi}~and~\eqref{eqn:sc_pij} and
chemical equilibrium.  This work was supported by the
European Research Council (ERC) Advanced Grant 227758, the Wolfson Merit
Award 2007/R3 of the Royal Society of London and the Engineering and
Physical Sciences Research Council (EPSRC) Programme Grant EP/I001352/1.
P.V. has been supported by a Marie Curie International Incoming Fellowship
of the European Community's Seventh Framework Programme under contract
number PIIF-GA-2011-300045, and by a Tizard Junior Research Fellowship from
Churchill College.
\end{acknowledgements}


\appendix

\section{Entropic cost of ligand binding}
\label{app:entropic}

In this appendix we describe how to calculate the free-energy change $\DGij$
of binding two ligands, $i$~and~$j$, when the binding interaction is
physically limited to a small, reactive sticky end.  We begin with a
general treatment, and later specialize to the two important cases of short,
stiff polymers and of freely-jointed chains.  These two models have been
used to describe the DNA strands on DNA-coated colloids when these are
short, rigid dsDNA~\cite{DreyfusEtAl:2009, LeunissenFrenkel:2011,
  MognettiLeunissenFrenkel:2012, Angioletti-UbertiMognettiFrenkel:2012} and
when they are long, flexible ssDNA~\cite{RogersCrocker:2011,
  MognettiEtAl:2012}.  These models may also be useful to describe
ligand-mediated binding in important biological systems, such as cells,
where adhesion is, in many cases, mediated by different glycoproteins coated
on their surface membrane\cite{BellDemboBongrand:1984}.

Let $Q_i$, $Q_j$ and $Q_{ij}$ be the partition functions of the two unbound
ligands and the bound complex that they form.  In terms of these quantities,
$\DGij$ is given by
\begin{equation}
\exp(-\beta\DGij) = \frac{Q_{ij}}{Q_i Q_j}\,.
\label{EqA1}
\end{equation}
As in the main text, we have assumed that the unbound ligands do not
otherwise interact with each other.  The Boltzmann factor
$\exp(-\beta\DGij)$ is generally smaller than that of the sticky ends in
solution at standard conditions, $\exp(-\beta\DGij^0)$, owing to the
entropic penalty of constraining the configurations available to ligands
$i$~and~$j$ when their reactive sticky ends are bound.  This defines the
configurational free energy cost of binding, $\DGij^\cnf$, as follows:
\begin{equation}
\DGij^\cnf = \DGij - \DGij^0.
\label{EqA2}
\end{equation}

A realistic model that reproduces $\DGij$ for ligands such as DNA strands is
expensive, as one would need to at least model each nucleotide of each
strand explicitly~\cite{OuldridgeLouisDoye:2011}.  It is computationally more convenient
to separate the binding of the sticky ends from the configurational freedom
of the rest of the ligand, and use a coarse-grained representation of the
sticky ends where $\DGij^0$ is introduced as an explicit external parameter.
Concretely, the internal partition functions of the free and bound sticky
ends in solution, $\cZ_i$, $\cZ_j$ and $\cZ_{ij}$, are related to the
equilibrium constant of binding~\cite{LeunissenEtAl:2010},
\begin{equation}
K = \frac{\exp(-\beta \DGij^0)}{\rho_0} = \frac{\cZ_{ij}}{\cZ_i \cZ_j},
\end{equation}
where $\rho_0$ is the standard concentration.  Using this relation in
Eq.~\eqref{EqA1} to factorize the unknown dependence of $Q_{ij}$,
$Q_i$~and~$Q_j$ on the internal partition functions, we obtain
\begin{equation}
\frac{Q_{ij}}{Q_i Q_j} \approx
\frac{\cZ_{ij}}{\cZ_i \cZ_j} \frac{Q^\cnf_{ij}}{Q^\cnf_i Q^\cnf_j},
\end{equation}
where $Q^\cnf_{ij}$, $Q^\cnf_i$~and~$Q^\cnf_j$ are approximately independent
of the chemical identity of the sticky end (``cnf'' stands for
``configurational'').  It follows that
\begin{equation}
\exp[-\beta\DGij^\cnf] \approx \frac{1}{\rho_0}
\frac{Q^\cnf_{ij}}{Q^\cnf_i Q^\cnf_j}.
\label{eqn:DGijcnf}
\end{equation}
We stress that this factorization is an approximation that holds only when
the binding of the sticky ends is only weakly affected by the presence of
the rest of the ligand, as occurs when the ligand is an ideal polymer or
when the binding is very strong.  However, it is a useful approximation
because once $\DGij^\cnf$ has been measured or calculated, one can predict
the values of $\DGij$ for families of ligands that differ only in their
sticky ends, providing us with a powerful design tool.

We now specialize to the case of polymeric ligands with one end fixed and
the other end sticky, such as polymers grafted at a surface.  The main
complication in treating polymers is dealing with excluded volumes.  For
instance, in DNACCs, the DNA strands cannot penetrate the colloids.  In the
discussion below, we denote by~$\vr_i$ the grafting point of
polymer~$i$. Given two polymers $i$~and~$j$, we denote by~$\vr_{ij}$ the
vector $\vr_j - \vr_i$ with length~$r_{ij}$.  We denote by $\Omega$ the
region of space that can be explored by the monomers of the polymers.
Polymers restricted to $\Omega$ are deemed ``confined'', whereas those for
which this restriction is absent are ``unconfined''.

We note that the free energy of repulsion due to a polymer~$i$, denoted by
$F_{\rep,i}$, is captured by the dependence of $Q_i^\cnf$ on the confinement
region~$\Omega$.  Indeed, letting ${\hat Q}_i^\cnf$ be the partition
function of polymer~$i$ in the absence of confinement, we have
\begin{equation}
e^{-\beta F_{\rep,i}} = Q_i^\cnf / {\hat Q}_i^\cnf.
\label{eqn:FRepDefn}
\end{equation}
From this expression follows the total free energy of repulsion,
\begin{equation}
F_\rep = \sum_i F_{\rep, i}.
\end{equation}
Similarly, the mean-field free energy density of repulsion is given by
\begin{equation}
f_\rep = \sum_a \sigma_a F_{\rep,a},
\label{eqn:frepMFT}
\end{equation}
where $\sigma_a$ is the coverage density of species~$a$ and $F_{\rep,a}$ is
the free energy of repulsion of any representative $a$-type strand. 

\subsection{Short, stiff polymers as ligands}
\label{app:entropic:rods}

A short, stiff polymer, such as a short piece of dsDNA, can be modeled as a
rigid rod, as has been done previously in Refs.~\onlinecite{LeunissenFrenkel:2011,
  MognettiLeunissenFrenkel:2012, Angioletti-UbertiMognettiFrenkel:2012}.  We
gather the results here for completeness and to clarify the origin of a
Jacobian prefactor absent in Ref.~\onlinecite{LeunissenFrenkel:2011} but addressed
in Ref.~\onlinecite{MognettiLeunissenFrenkel:2012}.  For rigid rods, the
configurational partition functions are:
\begin{subequations}
\begin{align}
Q^\cnf_{ij}&=\int_\Omega \dee\vr\,
\delta(|\vr-\vr_i| - L_i)
\delta(|\vr-\vr_j| - L_j), \label{eq:3a}\\
Q^\cnf_i&= \int_\Omega \dee\vr\, \delta(|\vr-\vr_i| - L_i), \label{eq:3b}\\
Q^\cnf_j&= \int_\Omega \dee\vr\, \delta(|\vr-\vr_j| - L_j), \label{eq:3c}
\end{align}
\end{subequations}
where $L_i$~and~$L_j$ are the lengths of rods $i$~and~$j$.

The first integral is most easily evaluated in cylindrical coordinates, with
the $z$~axis parallel to $\vr_{ij}$.  Special care
must be taken to isolate the Jacobian factor that arises when dealing with
the product of the two $\delta$-functions.  After evaluating the integrals,
we obtain,
\begin{align}
Q^\cnf_{ij}&=
\frac{\Omega_{ij}}{\sin(\gamma_i + \gamma_j)},\\
Q^\cnf_i&=\Omega_i,\\
Q^\cnf_j&=\Omega_j.
\label{eq:conf-space}
\end{align}
Here, $\Omega_{ij}$ is the portion within $\Omega$ of the circumference of
the circle explored by the common endpoint of the hybridized rods,
$\Omega_i$~and~$\Omega_j$ are the areas within~$\Omega$ of the sphere that
the endpoints of the free rods can explore, and $\gamma_i$ and $\gamma_j$
are the angles that the hybridized rods make with~$\vr_{ij}$.  The factor of
$1/\sin(\gamma_i + \gamma_j)$ was obtained in
Ref.~\onlinecite{MognettiLeunissenFrenkel:2012} using two different
representations of the $\delta$-function, but the procedure here highlights
the generality of the result.

When the colloids are plates, one at $z=0$ and one at $z=h$, the region
$\Omega$ is simply that where $0 \leq z \leq h$, and one can write an
explicit expression for $\Omega_{ij}$.  For the case where the rods are
grafted on opposite plates, such an expression is given in Appendix~A of
Ref.~\onlinecite{MognettiLeunissenFrenkel:2012} (see also Figure~8 therein).
When the rods are grafted to the same plate, $\Omega_{ij}$ is zero unless
$|L_i - L_j| \leq r_{ij} \leq L_i + L_j$, in which case
\begin{equation}
\frac{\Omega_{ij}}{L_i\sin\gamma_i} =
\begin{cases}
 \pi,&h \geq L_i\sin\gamma_i,\\
 2 \arcsin[h/(L_i\sin\gamma_i)],&h < L_i\sin\gamma_i.
\end{cases}
\end{equation}
The quantity $\Omega_i$ also has a compact expression,
\begin{equation}
\Omega_i =
\begin{cases}
2\pi L_i^2,&h \geq L_i,\\
2\pi L_i h,&0 < h \leq L_i.
\end{cases}
\end{equation}
The free energy of repulsion of a strand~$i$ follows immediately from
Eq.~\eqref{eqn:FRepDefn},
\begin{equation}
F_{\rep,i} = 
\begin{cases}
-\kT \ln(h/L_i),&0 < h \leq L_i,\\
0,&h \geq L_i.
\end{cases}
\label{eqn:FrepiRods}
\end{equation}

\subsection{Long, flexible polymers as ligands}
\label{app:entropic:polymers}

Following Refs.~\onlinecite{RogersCrocker:2011, MognettiEtAl:2012}, we can model a
polymer~$i$ as a freely-jointed chain (FJC) with $N_i$~segments of
length~$\ell$.  Let $\vr_{i,s}$ be the endpoint of segment~$s$, with $1 \leq
s \leq N$.  For convenience, let $\vr_{i,0} = \vr_i$, and let $\vu_{i,s} =
\vr_{i,s}-\vr_{i,s-1}$.  The chain segments are confined to the
region~$\Omega$, but do not otherwise interact.  It follows that
\begin{equation}
Q^\cnf_i = \int_\Omega \, \prod_{s=1}^{N_i} \dee \vu_{i,s}\,
\delta(|\vu_{i,s}| - \ell)\, ,
\label{EqA5}
\end{equation}
where, abusing notation somewhat, the integration domain confines the
polymer to the region~$\Omega$.  A completely analogous treatment yields
$Q^\cnf_j$ for a FJC with $N_j$ segments of length~$\ell$, grafted at
$\vr_j$.

One way of modeling the binding of $i$~and~$j$ is to coalesce the endpoints
of the two chains, so that $\vr_{i,N_i} = \vr_{j,N_j}$, resulting in a FJC
with endpoints fixed at $\vr_i$~and~$\vr_j$ and with $N_\hyb = N_i + N_j$
segments~$\{\vu_s\}$, where $1 \leq s \leq N_\hyb$.  However, if the
persistence length of the polymers is short, then the binding site may have
a size comparable to the segment length~$\ell$.  This is the case for ssDNA
polymers with sticky ends that are 6~to~12 nucleotides long. Hence, we also
consider a more general model for binding of $i$~and~$j$ in which the final
segment of the two polymers coalesces, so that $\vr_{i,N_i} = \vr_{j,N_j-1}$
and $\vr_{i,N_i-1} = \vr_{j,N_j}$, resulting in a FJC of $N_\hyb = N_i + N_j
- 1$ segments~\footnote{One can imagine coalescing more segments than the
  final one.  However, since the bound site is presumably poorly modeled by
  a FJC of the same persistence length as the unbound polymers, one would
  have to go beyond the treatment given here.}. The partition function of
the bound system is
\begin{align}
Q^\cnf_{ij} &= (4 \pi \ell^2)^{N_i + N_j - N_{\text{hyb}}}\, \tilde Q^\cnf_{ij},
\label{eqn:long_Qijcnf}
\\
\tilde Q^\cnf_{ij} &= \int_\Omega\,
\Bigl[\prod_{s=1}^{N_\hyb}\dee\vu_s\,\delta(|\vu_s|-\ell)\Bigr]
\delta\Bigl(\vr_{ij} - \sum_{s=1}^{N_\hyb} \vu_s\Bigr)\, ,
\label{EqA6}
\end{align}
where $\tilde Q^\cnf_{ij}$ is the partition function of a confined FJC with
fixed end points, $\vr_i$~and~$\vr_j$.  The additional factor in
$Q^\cnf_{ij}$ results from fusing the end segments of the individual
polymers.

Although the integrals in Eqs.~\eqref{EqA5}~and~\eqref{EqA6} cannot be
evaluated analytically even in simple geometries, as was done in the
previous section, one can calculate their value numerically using standard
polymer techniques.  Let $Q_1(N) = (4\pi\ell)^N$ be the partition function
of an unconfined $N$-segment FJC fixed at one endpoint, and let $Q_2(N,R)$
be the partition function of the same FJC with fixed end-to-end
distance~$R$.  Their ratio yields the probability density for the end-to-end
distance of the FJC to be~$R$,
\begin{equation}
p(N,R) = \frac{Q_2(N,R)}{Q_1(N)},
\label{eqn:Yamakawa}
\end{equation}
which is known in closed form~\cite{Treloar:1946, Yamakawa:1971}.  Using these partition
functions, we can write $\tilde Q^\cnf_{ij}$ in the following convenient
form:
\begin{equation}
\tilde Q^\cnf_{ij} = \frac{\tilde Q^\cnf_{ij}}{Q_2(N_\hyb, r_{ij})}
Q_2(N_\hyb, r_{ij}).
\label{eqn:preRosenbluth}
\end{equation}

The ratio on the right-hand side of Eq.~\eqref{eqn:preRosenbluth} is the
probability that a random walk from $\vr_i$~to~$\vr_j$ lies wholly
within~$\Omega$ (i.e., does not penetrate the colloid).  To compute it, we
use Rosenbluth sampling~\cite{FrenkelSmit:2001}, a common technique to
generate a properly weighted ensemble of such FJCs.  Specifically, to grow
each segment~$s$ of one chain that starts at $\vr_i$, we first generate $K$
possible segments $\vu^{(k)}_{s+1}$ ($1 \leq k \leq K$), sampled with
probability density $p(N_\hyb-(s+1),|\vr_j-\vr_{s+1}|)$ using rejection
sampling~\cite{Liu:2002}. We then pick at random one of the $m_s$ segments
in $\{\vu^{(k)}_{s+1}\}$ wholly within~$\Omega$.  The Rosenbluth weight of
that chain is defined as
\begin{equation}
W = \prod_{s=1}^{N_\hyb} \frac{m_s}{K}.
\label{eqn:Rosenbluth}
\end{equation}
The fraction of FJCs that lie wholly within~$\Omega$ is then the average
Rosenbluth factor,
\begin{equation}
\frac{\tilde Q^\cnf_{ij}}{Q_2(N_\hyb, r_{ij})} = \langle W \rangle_{ij},
\label{eqn:long_relQcnfij}
\end{equation}
where the subscript implies that the average is computed as described above.

We can estimate $Q^\cnf_i$~and~$Q^\cnf_j$ analogously, except that at each
step in growing the chain, now of length $N_i$ or $N_j$, the possible
segments $\{\vu^{(k)}_{s+1}\}$ are chosen uniformly on the surface of a
sphere of radius~$\ell$, since the end of the chain is free. We obtain
\begin{align}
\frac{Q^\cnf_i}{Q_1(N_i)} &= \langle W \rangle_i,\label{eqn:long_Qicnf}\\
\frac{Q^\cnf_j}{Q_1(N_j)} &= \langle W \rangle_j.\label{eqn:long_Qjcnf}
\end{align}

Substituting Eqs.~\eqref{eqn:long_Qijcnf}--\eqref{eqn:long_Qjcnf} into
Eq.~\eqref{eqn:DGijcnf}, we obtain the following expression for
the configurational entropy cost of binding $i$~and~$j$:
\begin{equation}
\exp(-\beta\DGij^\cnf) =
\frac{p(N_\hyb, r_{ij})}{\rho_0}
\frac{\langle W \rangle_{ij}}{\langle W \rangle_i \langle W \rangle_j}.
\label{eqn:DGcnfFJC}
\end{equation}

\begin{figure}
\begin{center}
\includegraphics{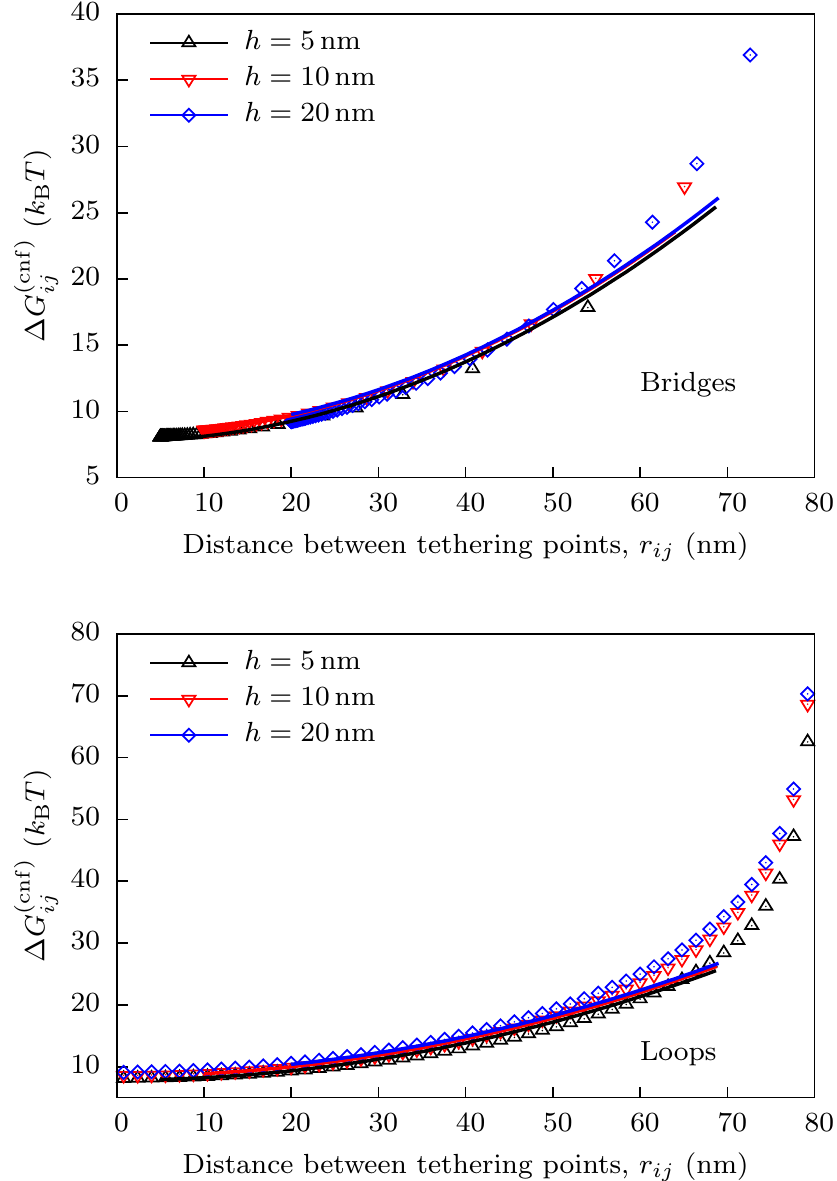}
\end{center}
\caption{\label{figDGcon}$\DGij^\cnf$ for a plane geometry and the polymeric
  constructs of Sec.~\ref{sec:results:ssDNA} as obtained using MC
  simulations (symbols) and a harmonic spring approximation
  (Eqs. \eqref{eqn:QiHarmonic},
  \eqref{eqn:QijHarmonicBridge}~and~\eqref{eqn:QijHarmonicLoop}, full
  lines). The two FJCs are tethered either to different plates (above)
  or to the same plate (below).}
\end{figure}

In Figure~\ref{figDGcon} we show the configurational entropy cost of
hybridizing two FJCs (made of $8$~segments with $\ell=5\,$nm) tethered to
parallel plates.  Interestingly, if $\DGij^\cnf$ is expressed in terms of the
distance between tethering points, $r_{ij}$, and the plate separation, $h$,
the dependence on the latter is marginal.  This suggests that, for ideal
chains, the cost of stretching a chain ($p(N_\hyb,r_{ij})$ in
Eq.~\eqref{eqn:DGcnfFJC}) is more important than the excluded volume terms
($W$). Hence, we may approximate $\DGij^\cnf$ analytically by neglecting all
chain degrees of freedom but the sticky end. In particular, using a Gaussian
approximation for $p(N,r)$~\cite{Yamakawa:1971},
\begin{equation}
p(N,r) \approx \left(\frac{3}{2\pi N \ell^2}\right)^{3/2}
               \exp\left[\frac{-3 r^2}{2N \ell^2}\right]
\end{equation}
we can model the ligand as a reactive point-like sticky end that is confined
between the two walls, and is attached to the tethering point via a harmonic
spring with spring constant~$k=3\,\kT/N\ell^2$. The partition function of an
unhybridized polymer is then given by
\begin{equation}
Q_i^\cnf \approx \int_{\Omega} \dee\vr \, p(N, |\vr - \vr_i|) = \frac12 \erf 
\left[ h \sqrt{\frac{3}{2 N \ell^2}} \right],
\label{eqn:QiHarmonic}
\end{equation}
where the integration domain, $\Omega$, is $[-\infty:\infty]^2\times[0:h]$,
and the error function, $\erf(x)$, is defined as in
Eq.~\eqref{eqn:erf_defn}.  Similarly, the partition function of the
hybridized polymer is given
\begin{equation}
Q^\cnf_{ij} \approx \int_{\Omega} \dee\vr\, p(N, |\vr - \vr_i|) \cdot p(N, |\vr
- \vr_j|).
\end{equation}
When $i$~and~$j$ are tethered to opposite plates, so that the bound pair
forms a bridge, we obtain
\begin{equation}
Q^{\mathrm{(bridge)}}_{ij} \approx
\frac{1}{8} \left( \frac{3}{\pi N\ell^2} \right)^{3/2}
  \erf\left(\frac{h}{2} \sqrt{\frac{3}{N\ell^2}}\right)
  \exp\left[ -\frac{3 r_{ij}^2}{4 N \ell^2}\right],
\label{eqn:QijHarmonicBridge}
\end{equation}
whereas when $i$~and~$j$ are tethered to the same plate, so that the bound
pair forms a loop, we obtain
\begin{equation}
Q^{\mathrm{(loop)}}_{ij} \approx
\frac{1}{16} \left(\frac{3}{\pi N\ell^2}\right)^{3/2}
  \erf\left(h \sqrt{\frac{3}{N\ell^2}}\right)
  \exp\left[ -\frac{3 r_{ij}^2}{4 N \ell^2}\right].
\label{eqn:QijHarmonicLoop}
\end{equation}
From these formulas, using Eq.~\eqref{eqn:DGijcnf}, we obtain analytical
estimates of $\DGij^\cnf$.  In Figure~\ref{figDGcon}, these estimates are
shown in the solid curve.  We see an overall agreement with the exact result
up to $r_{ij}=50\,$nm, where the finite extensibility of the chain becomes
important.

The above formulas can be used to derive an approximate expression for the
coefficients $K_{ab}$ used in the mean-field treatment of
Section~\ref{sec:theory:meanfield} (Eq.~\eqref{eqn:simpleKabFJC}).  Indeed,
for bridges,
\begin{equation}
K_{ab} \approx \int_h^\infty\dee r_{ab}\, 2\pi r_{ab}
\frac{Q^{\mathrm{(bridge)}}_{ab}}{Q^\cnf_a Q^\cnf_b},
\end{equation}
and for loops,
\begin{equation}
K_{ab} \approx \int_0^\infty \dee r_{ab}\, 2\pi r_{ab}
\frac{Q^{\mathrm{(loop)}}_{ab}}{Q^\cnf_a Q^\cnf_b}.
\end{equation}
In both of these cases, we have assumed that $Q^\cnf_{ab}$ goes to zero
sufficiently quickly for the integrals to be insensitive to their upper
bounds.

Finally, the above formulas also yield simple approximate expressions for
the free energy of repulsion, $F_{\rep,i}$, of a strand~$i$.  From
Eq.~\eqref{eqn:FRepDefn}, we obtain
\begin{equation}
F_{\rep,i} \approx -\kT\ln\erf\Bigl[ \sqrt{\frac{3 h^2}{2 N \ell^2}} \Bigr].
\label{eqn:FrepiFJC}
\end{equation}


\section{Relation to Local Chemical Equilibrium treatment}
\label{app:lce}

The derivation of Eqs. \eqref{eqn:sc_pi}~and~\eqref{eqn:sc_pij} can be
regarded as a ``ligand-centric'' result: the focus is solely on which
ligands (DNA strands) bind to one another.  We can take a complementary
``sticky-end-centric'' approach to obtain the same results, which sheds
light on the recent Local Chemical Equilibrium treatment of Rogers and
Crocker~\cite{RogersCrocker:2011} and corrects its main deficiency.

As in the main text, we first treat all strands separately, then make
continuum approximations for simplified geometries.  Following the notation
of Ref.~\onlinecite{RogersCrocker:2011}, let $C_i^0(\vr)$ be the normalized
probability density of finding the sticky end of tether~$i$ at~$\vr$, before
any hybridization occurs.  For this treatment only, we assume that the
strands interact ``ideally'', i.e., with an interaction energy $\Delta
G^0_{ij}$ when their sticky ends coincide and none otherwise, so that the
configurations of the tether of $i$ with sticky end at $\vr$ have the same
statistical weight whether the tethers are bound or unbound.  In terms of
partition functions, strands $i$~and~$j$ interact ideally if their
partition function when bound, $Q_{ij}$, is given by
\begin{equation}
Q_{ij} = \int\dee\vr\, Q_i(\vr) Q_j(\vr) \frac{\exp(-\beta\DG^0_{ij})}{\rho_0},
\end{equation}
where $Q_i(\vr)$ is the partition function of unbound strand~$i$ with sticky
end at $\vr$, and $Q_j(\vr)$ is defined similarly.  The restriction of ideal
binding, absent in the general treatment of the main text, is satisfied when
the tethers are ideal chains, for example.  With this additional assumption,
we rewrite Eq.~\eqref{eqn:Z} as follows:
\begin{equation}
\frac{Z}{Z_0} = \sum_{\phi} \prod_{(i,j) \in \phi} \int\dee\vr\, C_i^0(\vr)
C_j^0(\vr) \frac{\exp(-\beta\DG^0_{ij})}{\rho_0},
\end{equation}
where $\rho_0$ is the standard concentration, $1\,$M.  The analog of
Eq.~\eqref{eqn:sc_Z} is:
\begin{multline}
1 = \int\dee\vr\, \Biggl\{\frac{Z_{-i}}{Z} C_i^0(\vr) \\
+ \sum_j C_i^0(\vr)
C_j^0(\vr) \frac{\exp(-\beta\DG^0_{ij})}{\rho_0} \frac{Z_{-i,-j}}{Z}\Biggr\}.
\end{multline}
As in the main text, let $p_i = Z_{-i} / Z$, and approximate $Z_{-i,-j}/Z$
by~$p_i p_j$.  As in Ref.~\onlinecite{RogersCrocker:2011}, let $C_i(\vr)$ be the
concentration of free sticky end~$i$ at~$\vr$, equal to $p_i C_i^0(\vr)$ by
comparison to the equation above.  There being only one such sticky end,
$C_i(\vr)$ is also the probability density that the sticky end of tether~$i$
is free and is at~$\vr$.  We obtain
\begin{equation}
1 = \int\dee\vr\, \left\{C_i(\vr) + \sum_j C_i(\vr)
C_j(\vr) \frac{\exp(-\beta\DG^0_{ij})}{\rho_0}\right\}.
\end{equation}
More concisely, let $C_{ij}(\vr)$ be the concentration of bound sticky ends
$i$~and~$j$, given by
\begin{equation}
C_{ij}(\vr)
= C_i(\vr) C_j(\vr) \frac{\exp(-\beta\DG^0_{ij})}{\rho_0}.
\end{equation}
We can then finally write the self-consistent condition for the
quantities~$\{p_i\}$ as follows:
\begin{subequations}
\begin{align}
1 &= \int\dee\vr\, \left\{ C_i(\vr) + \sum_j C_{ij}(\vr) \right\},\\
C_i(\vr) &= p_i C_i^0(\vr).
\end{align}
\label{eqn:CorrectLCE}
\end{subequations}

Comparing the above condition to the equations in
Ref.~\onlinecite{RogersCrocker:2011}, we conclude that the ``local chemical
equilibrium'' approximation introduced by the authors is equivalent to our
approximation $Z_{-i,-j}/Z \approx p_i p_j$, but that the self-consistent
condition is significantly different from the condition that they proposed,
\begin{equation}
C_i(\vr) \stackrel{?}{=} C_i^0(\vr) - \sum_j C_{ij}(\vr).
\label{eqn:RogersCrocker}
\end{equation}
Physically, in the treatment of Rogers and Crocker, depletion of free sticky
ends~$i$ at a point~$\vr$ establishes a density gradient in $C_i(\vr)$
beyond that implied by the polymer statistics of tether~$i$, but this
density gradient is not allowed to relax, effectively imposing a
nonequilibrium condition.  Our treatment, summarized in
Eq.~\eqref{eqn:CorrectLCE}, corrects this shortcoming.

As in the main text, for those cases where there is translational
invariance, we may approximately treat together tethers with similar binding
behavior.  In the concrete case of two plates perpendicular to the $z$~axis,
and in the absence of binding, the concentration of $a$-type sticky ends at
a height~$z$ is:
\begin{equation}
C_a^0(z) = \sigma_a P_a(z),
\end{equation}
where
\begin{equation}
P_a(z) = \int\dee x\,\dee y\, C_i^0(x,y,z),
\end{equation}
and $i$ is any representative tether of type~$a$.  The corresponding mean
field self-consistent equations are
\begin{align}
\sigma_a &= \int\dee z\, \left\{ C_a(z) + \sum_b C_{ab}(z) \right\},\\
C_a(z) &= p_a C_a^0(z),\\
C_{ab}(z) &= C_a(z) C_b(z) \frac{\exp(-\beta\DG^0_{ab})}{\rho_0}.
\end{align}
Rearranging these equations to obtain explicit equations for $\{p_a\}$
yields
\begin{equation}
p_a = \frac{1}{1 + p_b \sigma_b \int\dee z\,P_a(z) P_b(z) \frac{\exp(-\beta\DG^0_{ab})}{\rho_0}}
\end{equation}
Comparing this last equation to Eq.~\eqref{eqn:scmf_pa}, we find that
\begin{equation}
K_{ab} = \frac{e^{-\beta\DG^0_{ab}}}{\rho_0} \int\dee z\, P_a(z) P_b(z).
\end{equation}
This is a specialization of Eq.~\eqref{eqn:scmf_Kab} under the additional
assumption of ideal binding interactions.

For the concrete case of short, rigid-rod-like tethers of length~$L$, the
probabilities~$P_a(z)$ are particularly simple, and depend only on rod
length, plate separation, and the plate on which $a$-type tethers are
grafted. In particular, if they are grafted on the plate at $z=0$ and the
other plate is at $z=h$, then
\begin{equation}
P_a(z) = \begin{cases}
1 / h,& 0 < h < L;\\
1 / L,& h > L \text{ and } 0 < z < L;\\
0,&\text{otherwise}.
\end{cases}
\end{equation}
Otherwise, if $a$-type tethers are grafted on the $z=h$ plate, then
\begin{equation}
P_a(z) = \begin{cases}
1 / h,& 0 < h < L;\\
1 / L,& h > L \text{ and } h-L < z < L;\\
0,& \text{otherwise}.
\end{cases}
\end{equation}
From these results, one can easily obtain Eq.~\eqref{eqn:simpleKab}.  This
derivation should be compared with the explicit integrals involved in
calculating $K_{ab}$ using Eq.~\eqref{eqn:scmf_Kab} and the relevant
expressions for $\DGij^\cnf$ given in Appendix~\ref{app:entropic} of this
paper and Appendix~A of Ref.~\onlinecite{MognettiLeunissenFrenkel:2012}.  We
have verified that both routes give identical results when evaluated
numerically for $L_b = L_a$ and $L_b = 1.7 L_a$.

In the absence of translational invariance or some similar symmetry, it is
nontrivial, in general, to write down self-consistency conditions similar to
Eq.~\eqref{eqn:CorrectLCE} where like-type tethers are grouped together.
Unlike the incorrect condition of Eq.~\eqref{eqn:RogersCrocker}, the correct
self-consistency conditions are inherently non-local from the point of view
of sticky ends: the change upon binding in sticky end concentration of
$a$-type sticky ends at $\vr$ depends on the \emph{precise} identity of the
$a$-type tether to which that sticky end is attached.  Indeed, one needs to
retain the level of detail inherent in Eq.~\eqref{eqn:CorrectLCE}.  As such,
there is no advantage in adopting a ``sticky-end-centric'' point of view as
opposed to a ``tether-centric'' point of view when dealing with arbitrarily
shaped colloids, polymer statistics or non-uniform coverage.  Moreover, the
tether-centric equations are simpler, since they result from integrating
over the irrelevant degrees of freedom describing the position of the sticky
ends, and are more general, since they do not assume that binding is ideal.

To conclude this section, we present a numerical comparison between Monte
Carlo results, our self-consistent theory and the LCE theory that
illustrates the extent to which these approximate theories deviate from an
exact treatment.  We consider two periodic parallel plates, each with $500$
tethers, at a density comparable to that used in
Ref.~\onlinecite{RogersCrocker:2011} ($4500/(4 \pi \times 550^2) =
0.001183\,$tethers/nm$^2$).  The tethers are modeled as ideal $8$-segment
chains of Kuhn length $5\,$nm that, when bound, form a $16$-segment ideal
chain, as described in Appendix~\ref{app:entropic}.
Figure~\ref{fig:ComparisonToLCE} shows the number of bonds formed in this
system as a function of the solution hybridization free energy, $\Delta
G^0$, of the sticky ends of tethers on opposite plates, when the plates are
separated by $15\,$nm and by $25\,$nm.  As is observed, our self-consistent
theory closely tracks the Monte Carlo results, whereas the estimates
obtained using the LCE approach of Rogers and Crocker differ appreciably
from the exact ones.

\begin{figure}
\begin{center}
\includegraphics{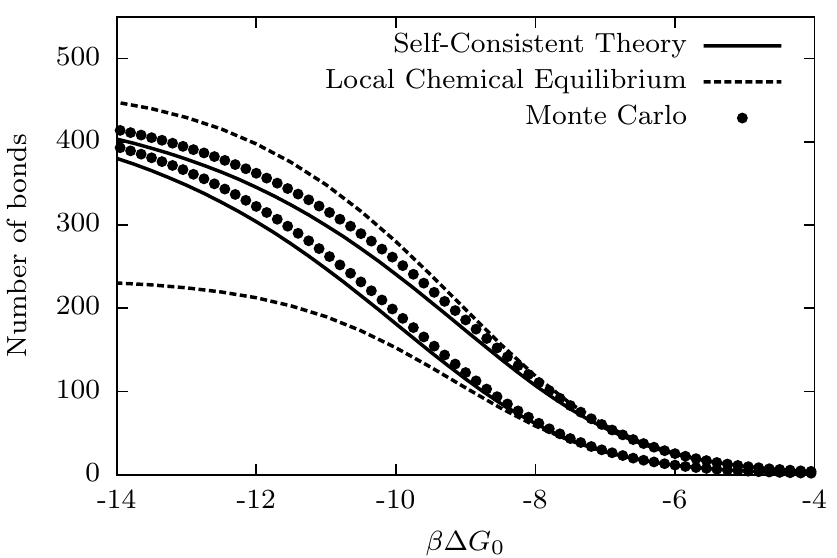}
\end{center}
\caption{\label{fig:ComparisonToLCE}Exact and estimated number of bonds
  formed between two ssDNA-coated parallel plates separated by $15\,$nm
  (lower curves) and $25\,$nm (upper curves).  The Self-Consistent Theory is
  that of Eqs.  \eqref{eqn:sc_pi}~and~\eqref{eqn:sc_pij}, while the Local
  Chemical Equilibrium curve results from the methods of
  Ref.~\onlinecite{RogersCrocker:2011}.  See text for details.}
\end{figure}


\section{Self-consistent equations from a Saddle Point Approximation}
\label{app:saddlepoint}

In this appendix, we present an alternate derivation for the self-consistent
mean field theory in Eqs. \eqref{eqn:scmf_pa}~and~\eqref{eqn:scmf_Kab}
in two specific setups, based on a saddle point approximation of a
mean-field partition function.  The two geometries we explore are the first
two examples in Section~\ref{sec:results}, which we refer to as the single
linkage system and the competing linkages system.

\subsection{Single linkage system}
\label{app:saddlepoint:single}

In the single linkage case, two parallel plates of area~$A$ are respectively
decorated with $N_a$~and~$N_b$ strands of two complementary kinds of DNA,
$a$~and~$b$, whose surface concentration is thus $\sigma_a = N_a/A$ and
$\sigma_{b} = N_b/A$. Without loss of generality, we assume that $N_a\leq
N_b$. Let $n_{b \from a}$ be the average number of $b$-type strands that a
fixed $a$-type strand can bind to.  When no bonds have yet formed $n_{b\from
  a} = \sigma_b A_{b\from a}$, where $A_{b\from a}$ is the area that
encloses all possible grafting points of $b$-type binding partners of a
fixed $a$-type strand.  However, after having formed $n_{ab}$ bonds between
$a$- and $b$-type strands the average number of available linkers is reduced
to~\cite{MognettiLeunissenFrenkel:2012}
\begin{equation}
n_{b\from a}(n_{ab}) = \frac{N_b - n_{ab}}{N_b} \sigma_b A_{b\from a} \, .
\label{SP:comb}
\end{equation}
The average Boltzmann factor for binding $a$- and $b$-type strands is
$K_{ab} / A_{b\from a}$, where $K_{ab}$ is defined in
Eq.~\eqref{eqn:scmf_Kab}.  With this results and the previous one, we can
write a mean-field partition function of the system as follows
\begin{align}
Z &= \sum_{n_{ab}=0}^{N_a} {N_a \choose n_{ab}} \Biggl[\prod_{j=0}^{n_{ab}-1}
  \frac{K_{ab}}{A_{b\from a}} n_{b\from a}(j) \Biggr]\, ,\notag\\
&= \sum_{n_{ab}=0}^{N_a}
{N_a \choose n_{ab}} {N_b \choose n_{ab}} \frac{n_{ab}!}{N_b^{n_{ab}}} \,
(K_{ab} \sigma_b)^{n_{ab}} \, .
\label{SP:Zsingle}
\end{align}

We now isolate the dependence of $Z$ on $A$ in order to take the limit
$A\to\infty$, which we then evaluate through a saddle-point approximation.
Let $s_{ab} = n_{ab}/A$, and approximate the sum in Eq.~\eqref{SP:Zsingle}
by an integral with respect to $s_{ab}$.  Assuming the maximum of the
integrand is far from the integration limits, we obtain
\begin{equation}
Z \approx A \int \dee s_{ab}\, \exp[A \cdot \cS(s_{ab})],
\end{equation}
where, using Stirling's approximation,
\begin{multline}
\cS(s) =
\sigma_a \ln \sigma_a + \sigma_b \ln \sigma_b
-s\ln s -s
\\
-(\sigma_a-s)\ln(\sigma_a-s)
-(\sigma_b-s)\ln(\sigma_b-s)
\\
+s\ln K_{ab}\, .
\end{multline}
In the limit $A\to\infty$, the partition function will be dominated by the
values of $s$ that maximize $\cS(s)$.  Setting $\dee\cS/\dee s$ to $0$ at $s =
\sigma_{ab}$, we obtain the following equation for the average density of
hybridized strands, $\sigma_{ab}$:
\begin{equation}
\sigma_{ab} = (\sigma_a - \sigma_{ab}) K_{ab} (\sigma_b - \sigma_{ab}),
\label{SP:pab}
\end{equation}
which restates Eq.~\eqref{eqn:scmf_pab} after identifying $\sigma_a
p_a$ with $\sigma_a - \sigma_{ab}$ and likewise for $\sigma_b p_b$.  Using
Eq.~\eqref{eqn:scmf_pa} one can also rewrite Eq.~\eqref{SP:pab} as
\begin{equation}
\sigma_{ab} =
\frac{\sigma_a K_{ab} (\sigma_b-\sigma_{ab})}
     {1+ K_{ab}(\sigma_b-\sigma_{ab})} \,,
\end{equation}
which is the expression used in Ref.~\onlinecite{MognettiLeunissenFrenkel:2012}.

\subsection{Competing linkages system}
\label{app:saddlepoint:competing}

In this section we generalize the saddle point approach to the competing
linkages system.  Like in Section~\ref{sec:results}, we have two
parallel plates of area $A$, one grafted with $a$- and $b$-type strands,
the other with $a'$- and $b'$-type strands.  The only allowed interactions
are $a$--$a'$, $a$--$b$ and $a'$--$b'$.  An equivalent treatment applies
when, as in Ref.~\onlinecite{MognettiLeunissenFrenkel:2012}, allowed interactions are $a$--$a'$,
$a$--$b'$ and $a'$--$b$, so that loops (intra-particle bonds) are forbidden.  The only differences between the two systems are the labels in the equations and the values of the $K_{ab}$ factors below.

Using the same notation as in the previous section, we find that the average
number of binding partners available to a fixed strand of each type is
\begin{subequations}
\begin{align}
n_{a' \from a}(n_{aa'},n_{a'b'}) &=
 \frac{N_{a'}-n_{aa'}-n_{a'b'}}{N_{a'}} \, \sigma_{a'} \, A_{a'\from a}  \, ,\\
n_{b \from a}(n_{ab}) &=
 \frac{N_{b}-n_{ab}}{N_{b}} \, \sigma_{b} \, A_{b\from a} \, ,\\
n_{b' \from a'}(n_{a'b'}) &=
 \frac{N_{b'}-n_{a'b'}}{N_{b'}} \, \sigma_{b'} \, A_{b' \from a'} \, .
\end{align}
\label{SP:EqCombComp}
\end{subequations}
Assume for simplicity that $N_a\leq N_{a'}$, $N_a \leq N_b$ and
$N_{a'}\leq N_{b'}$.  A nearly identical argument applies for a different
relative populations of strands.  A mean-field partition function for the
system is
\begin{widetext}
\begin{align}
Z &=
\sum_{n_{aa'}=0}^{N_a} \sum_{n_{ab}=0}^{N_a-n_{aa'}} \sum_{n_{a'b'}=0}^{N_{a'}-n_{aa'}}
{N_a\choose n_{aa'}}
\Biggl[\prod_{k=0}^{n_{aa'}-1} \frac{K_{aa'}}{A_{a'\from a}} n_{a'\from a}(k,0)\Biggr]
\nonumber\\
&\qquad\qquad
\cdot{N_a-n_{aa'}\choose n_{ab}}
\Biggl[\prod_{t=0}^{n_{ab}-1} \frac{K_{ab}}{A_{b\from a}} n_{b\from a}(t)\Biggr]
{N_{a'}-n_{aa'}\choose n_{a'b'}}
\Biggl[\prod_{v=0}^{n_{a'b'}-1} \frac{K_{a'b'}}{A_{b'\from a'}} n_{b'\from a'}(v)\Biggr]
\nonumber\\
&=
\sum_{n_{aa'}=0}^{N_a} \sum_{n_{ab}=0}^{N_a-n_{aa'}} \sum_{n_{a'b'}=0}^{N_{a'}-n_{aa'}}
{N_a\choose n_{aa'}}
{N_{a'}\choose n_{aa'}}{n_{aa'}!\over N_{a'}^{n_{aa'}}}
(K_{aa'}\sigma_{a'})^{n_{aa'}}
\nonumber\\
&\qquad\qquad
\cdot{N_a-n_{aa'}\choose n_{ab}}
{N_{b}\choose n_{ab}}{n_{ab}!\over N_{b}^{n_{ab}}}
(K_{ab}\sigma_{b})^{n_{ab}}
{N_{a'}-n_{aa'}\choose n_{a'b'}}
{N_{b'}\choose n_{a'b'}}{n_{a'b'}!\over N_{b'}^{n_{a'b'}}}
(K_{a'b'}\sigma_{b'})^{n_{a'b'}} \, .
\label{eqn:SP:competingZ}
\end{align}
\end{widetext}

As in the previous section, we isolate the dependence of $Z$ on $A$ to take
the $A\to\infty$ limit.  Let $s_{xy} = n_{xy}/A$, and approximate the sums
in Eq.~\eqref{eqn:SP:competingZ} by integrals with respect to $s_{aa'}$,
$s_{ab}$~and~$s_{a'b'}$.  Assuming the maximum of the integrand is far from
the integration limits, we obtain
\begin{equation}
Z \approx {A^3} \iiint \dee s_{aa'}\, \dee s_{ab}\, \dee s_{a'b'}\,
\exp[A \cdot \cS(s_{aa'}, s_{ab}, s_{a'b'})],
\end{equation}
where
\begin{multline}
\cS(s_{aa'}, s_{ab}, s_{a'b'}) = \text{const.}\\
+ s_{aa'} (\ln {K_{aa'}\over s_{aa'}}-1)
+ s_{ab} (\ln {K_{ab}\over s_{ab}}-1)
+ s_{a'b'} (\ln {K_{a'b'}\over s_{a'b'}}-1)\\
- (\sigma_b - s_{ab}) \ln[\sigma_b - s_{ab}]
- (\sigma_{b'} - s_{a'b'}) \ln[\sigma_{b'} - s_{a'b'}]\\
- (\sigma_a - s_{aa'} - s_{ab}) \ln[\sigma_a - s_{aa'} - s_{ab}]\\
- (\sigma_{a'} - s_{aa'} - s_{a'b'}) \ln[\sigma_{a'} - s_{aa'} - s_{a'b'}]
\, ,
\end{multline}
where the first, constant term does not depend on $s_{xy}$.

The point $(\sigma_{aa'}, \sigma_{ab}, \sigma_{a'b'})$ at which $\cS$ is
maximized, found by setting all its partial derivatives to zero, yields
the average bond densities for each pair of species.  They satisfy
\begin{subequations}
\begin{align}
\sigma_{aa'} &= (\sigma_{a}-\sigma_{aa'}-\sigma_{ab}) K_{aa'}
  (\sigma_{a'}-\sigma_{aa'}-\sigma_{a'b'})
\nonumber \\
& = \sigma_a p_a K_{aa'} p_{a'} \sigma_{a'},\label{SP:saddlecomp_a}\\
\sigma_{ab} &= (\sigma_{a}-\sigma_{aa'}-\sigma_{ab'}) K_{ab}
  (\sigma_{b}-\sigma_{ab})
\nonumber \\
& = \sigma_a p_a K_{ab} p_{b} \sigma_{b},\label{SP:saddlecomp_b}\\
\sigma_{a'b'} &= (\sigma_{a'}-\sigma_{aa'}-\sigma_{a'b'}) K_{a'b'}
  (\sigma_{b'}-\sigma_{a'b'})
\nonumber \\
& = \sigma_{a'} p_{a'} K_{a'b'} p_{b'} \sigma_{b'} \, .
\label{SP:saddlecomp_c}
\end{align}
\label{SP:saddlecomp}
\end{subequations}
\noindent These equations correspond to the mean-field self-consistent theory of
Eqs. \eqref{eqn:scmf_pa}~and~\eqref{eqn:scmf_pab}) for the competing
linkages model.

We now show that the saddle points equations \eqref{SP:saddlecomp} are also
equivalent to the expressions derived in
Ref.~\onlinecite{MognettiLeunissenFrenkel:2012} (once $\sigma_{ab}$ and
$\sigma_{a'b'}$ are replaced with $\sigma_{ab'}$ and $\sigma_{a'b}$
respectively).  Using Eq.~\eqref{SP:saddlecomp_c} two times,
we have
\begin{align}
&\sigma_{a'} - \sigma_{aa'}-\sigma_{a'b'}\nonumber\\
&\quad=
 \frac{\sigma_{a'} -\sigma_{aa'}}
   {(\sigma_{a'} -\sigma_{aa'} -\sigma_{a'b'}) + \sigma_{a'b'}}
\cdot
\frac{\sigma_{a'b'}}{K_{a'b'}(\sigma_{b'}-\sigma_{a'b'})}
\nonumber\\
&\quad= \frac{\sigma_{a'} -\sigma_{aa'}}{1+ K_{a'b'}(\sigma_{b'}-\sigma_{a'b'})} \, .
\end{align}
Substituting this result into Eq.~\eqref{SP:saddlecomp_a}, together 
with the definition of $p_a$ in Eq.~\eqref{eqn:scmf_pa}, we find that
\begin{align}
\sigma_{aa'} &=
\sigma_a K_{aa'}
\frac{\sigma_{a'} -\sigma_{aa'}}{1+K_{a'b'}(\sigma_{b'}-\sigma_{a'b'})}
\nonumber\\
&\qquad\cdot\Biggl[1+{K_{aa'}(\sigma_{a'}
    -\sigma_{aa'})\over1+K_{a'b'}(\sigma_{b'}-\sigma_{a'b'})}
+ K_{ab}(\sigma_{b}-\sigma_{ab})\Biggr]^{-1}
\nonumber\\
&= \sigma_a K_{aa'}(\sigma_{a'} -\sigma_{aa'}) \cdot \Bigl[
1+K_{a'b'}(\sigma_{b'}-\sigma_{a'b'})
\nonumber \\
&\qquad+ K_{aa'}(\sigma_{a'} -\sigma_{aa'})
+K_{ab}(\sigma_{b}-\sigma_{ab})
\nonumber \\
&\qquad+K_{ab}(\sigma_{b}-\sigma_{ab})\cdot K_{a'b'}(\sigma_{b'}-\sigma_{a'b'})\Bigr]^{-1}\,.
\label{eurSCMF}
\end{align}
This is exactly the expression used in
Ref.~\onlinecite{MognettiLeunissenFrenkel:2012} to compute the density of
$a$--$a'$ linkages, derived there by applying an heuristic self-consistent
prescription to an approximate estimate of $\sigma_{aa'}$.  A similar
argument recovers the expressions for $\sigma_{ab}$~and~$\sigma_{a'b'}$ used
in Ref.~\onlinecite{MognettiLeunissenFrenkel:2012}.

\bibliography{bibl}

\end{document}